\documentclass[12pt]{article}
\pdfoutput=1

\usepackage{amssymb}
\usepackage{amsmath}
\usepackage{amscd}
\usepackage{mathrsfs}
\usepackage{amssymb}
\usepackage{feynmf}
\usepackage[pdftex]{graphicx}
\usepackage{youngtab}

\usepackage{float}

\allowdisplaybreaks

\def\dj{\hbox{d\kern-0.347em \vrule width 0.3em height 1.252ex depth
-1.21ex \kern 0.051em}}

\numberwithin{equation}{section}

\begin{document}

\setlength{\oddsidemargin}{0cm}
\setlength{\baselineskip}{7mm}


\thispagestyle{empty}
\setcounter{page}{0}

\begin{flushright}

\end{flushright}

\vspace*{0.5cm}

\begin{center}
{\bf \Large Projectivity of Planar Zeros in Field and }

\vspace*{0.3cm}

{\bf \Large String Theory Amplitudes}

\vspace*{1cm}

Diego Medrano Jim\'enez$^{a,}$\footnote{\tt  d.medrano@csic.es}, 
Agust\'{\i}n Sabio Vera$^{a,}$\footnote{\tt 
a.sabio.vera@gmail.com}
and Miguel \'A. V\'azquez-Mozo$^{b,}$\footnote{\tt 
Miguel.Vazquez-Mozo@cern.ch}

\end{center}

\vspace*{0.0cm}

\begin{center}

$^{a}${\sl Instituto de F\'{\i}sica Te\'orica UAM/CSIC \&
Universidad Aut\'onoma de Madrid\\
C/ Nicol\'as Cabrera 15, E-28049 Madrid, Spain
}

$^{b}${\sl Departamento de F\'{\i}sica Fundamental,  Universidad de Salamanca \\ 
 Plaza de la Merced s/n,
 E-37008 Salamanca, Spain
  }
\end{center}

\vspace*{1.0cm}

\centerline{\bf \large Abstract}

\noindent

We study the projective properties of planar zeros of tree-level scattering amplitudes in various theories. 
Whereas for pure scalar field theories we find that the planar zeros of the five-point amplitude 
do not enjoy projective invariance, 
coupling scalars to gauge fields gives rise to tree-level amplitudes whose planar zeros
are determined by homogeneous polynomials in the stereographic coordinates labelling the 
direction of flight of the outgoing particles. In the case of pure gauge theories, this
projective structure is generically destroyed if string corrections are taken into account. 
Scattering amplitudes of two scalars with graviton emission vanish exactly in the planar limit, whereas
planar graviton amplitudes are zero for helicity violating configurations. These results are corrected by 
string effects, computed using the single-valued projection, 
which render the planar amplitude nonzero. Finally, we discuss how the structure of
planar zeros can be derived from the soft limit behavior of the scattering amplitudes.

\newpage  


\setcounter{footnote}{0}

\section{Introduction}

Zeros in scattering amplitudes are useful devices to test interesting properties of the standard model. 
For example, the vanishing of the
tree-level amplitude of certain processes involving the emission of a gauge boson is very sensitive to
the form of the trilinear couplings.
Thus, the detection of amplitude zeros were proposed as a way to constraint the existence of anomalous couplings in the 
standard model \cite{rad-zeros} (see \cite{rad_zeros_rev} for reviews). Although these so-called type-I zeros 
are corrected by both loops and higher-order emissions, they manifest themselves 
in the existence of dips for a set of observables, a fact that has been
confirmed by various experimental groups \cite{zeros_exp}.

A second class of amplitude zeros appear for particular kinematic configurations in which all momenta 
are confined to a plane \cite{heyssler_stirling}. 
The phenomenological implications of these planar (or type-II) zeros has been recently studied in \cite{harland-lang} in the context of a
five parton amplitude, and it was shown how the planar zeros are determined by simple relations involving rapidity differences.

In a previous paper \cite{ours_pz}, we have studied the mathematical structure of planar zeros in gauge theories and gravity, focusing
on the five-point amplitude for gluons and gravitons. There it was found 
that,  once the outgoing momenta are expressed in terms of stereographic coordinates, 
the loci of planar zeros is determined by a cubic integer 
curve in the projective plane defined by these coordinates. 

Although the analysis presented in \cite{ours_pz} focused on the five-point scattering amplitude, 
the projective nature of the planar zeros in Yang-Mills theories is present for any multiplicity. 
To see this let us recall that, in a (super) Yang-Mills theory, 
the $n$-gluon tree-level amplitude can be written in the form
\cite{delducca_et_al}
\begin{align}
A_{n}=(ig)^{n-2}\sum_{\sigma\in S_{n-2}}c_{\sigma}A_{n}\Big(1,2,\sigma(3\ldots,n)\Big),
\end{align} 
where $A_{n}(1,\ldots,n)$ is the color-ordered amplitude and the color factors $c_{\sigma}$ are defined in terms
of the structure contants by
\begin{align}
c_{\sigma}=f^{a_{1}a_{2}c_{1}}f^{c_{1}a_{\sigma(3)}c_{2}}\ldots f^{c_{n-3}a_{\sigma(n-1)}a_{\sigma(n)}}.
\end{align}
To evaluate the amplitude in the planar limit, it is convenient to consider 
a center-of-mass frame where the incoming particles propagate along the $z$ axis:
\begin{align}
p_{1}&={\sqrt{s}\over 2}(1,0,0,1),\nonumber \\[0.2cm] 
p_{2}&={\sqrt{s}\over 2}(1,0,0,-1),
\label{eq:momentum_parametrization_in}
\end{align}
while the momenta of the on-shell outgoing gluons can be parametrized in terms of stereographic coordinates according to
\begin{align}
p_{a}&=-\omega_{a}\left(1,{\zeta_{a}+\overline{\zeta}_{a}\over 1+\zeta_{a}\overline{\zeta}_{a}},i{\overline{\zeta}_{a}-\zeta_{a}\over
1+\zeta_{a}\overline{\zeta}_{a}},{\zeta_{a}\overline{\zeta}_{a}-1\over 1+\zeta_{a}\overline{\zeta}_{a}}\right) \hspace*{0.5cm}
\mbox{with}\hspace*{0.5cm} a=3,\ldots,n.
\label{eq:momentum_parametrization_out}
\end{align}

For the particular case of MHV amplitudes, the Parke-Taylor formula \cite{parke_taylor} gives the following
expression for the color-ordered amplitudes
\begin{align}
A_{n}\Big(1^{-},2^{-},\sigma(3^{+},\ldots,n^{+})\Big)=i{\langle 12\rangle^{4}\over 
\langle 12\rangle\langle 2\sigma(3)\rangle\ldots \langle \sigma(n-1)\sigma(n)\rangle\langle\sigma(n)1\rangle}.
\label{eq:parke-taylor_eq}
\end{align}
Without loss of generality, we can consider planar scattering where all momenta lie on the plane $y=0$, which means that
the stereographic coordinates giving the direction of flight of the outgoing gluons are all real ($\zeta_{a}=
\overline{\zeta}_{a}$). The relevant spinor inner products are computed to be
\begin{align}
\langle 12\rangle &= \sqrt{s}, \nonumber \\[0.2cm]
\langle 1\sigma(j)\rangle&=-i\sqrt{2}s^{1\over 4}{|\zeta_{\sigma(j)}|\over \zeta_{\sigma(j)}}
\sqrt{\omega_{\sigma(j)}\over 1+\zeta_{\sigma(j)}^{2}}, \nonumber \\[0.2cm]
\langle 2\sigma(j)\rangle&=-i\sqrt{2}s^{1\over 4}|\zeta_{\sigma(j)}|\sqrt{\omega_{\sigma(j)}\over 
1+\zeta_{\sigma(j)}^{2}}, \\[0.2cm]
\langle\sigma(j)\sigma(k)\rangle&=2{\zeta_{\sigma(j)}(\zeta_{\sigma(j)}-\zeta_{\sigma(k)})\zeta_{\sigma(k)}
\over |\zeta_{\sigma(j)}||\zeta_{\sigma(k)}|}\sqrt{\omega_{\sigma(j)}\omega_{\sigma(k)}\over
(1+\zeta_{\sigma(j)}^{2})(1+\zeta_{\sigma(k)}^{2})}.
\nonumber
\end{align}
Plugging these expressions into Eq. \eqref{eq:parke-taylor_eq}, we find the following structure for the
MHV amplitude
\begin{align}
A_{n}\Big(1^{-},2^{-},\sigma(3^{+},\ldots,n^{+})\Big)=is\left(\prod_{i=3}^{n}{1+\zeta_{i}^{2}
\over \omega_{i}}\right)f_{\sigma}(\zeta_{3},\ldots,\zeta_{n}),
\end{align} 
where $f_{\sigma}(\zeta_{3},\ldots,\zeta_{n})$ is a rational homogeneous function of degree $2-n$:
\begin{align}
f_{\sigma}(\lambda\zeta_{3},\ldots,\lambda\zeta_{n})=\lambda^{2-n}f_{\sigma}(\zeta_{3},\ldots,\zeta_{n}),
\end{align}
for any $\lambda\neq 0$. Thus, the color-dressed planar tree-level $n$-gluon amplitude reads
\begin{align}
A_{n}=is(ig)^{n-2}
\left(\prod_{i=3}^{n}{1+\zeta_{i}^{2}\over \omega_{i}}\right)\sum_{\sigma\in S_{n-2}}
c_{\sigma}f_{\sigma}(\zeta_{3},\ldots,\zeta_{n}),
\end{align}
and the planar zeros are determined by the homogeneous equation
\begin{align}
\sum_{\sigma\in S_{n-2}}
c_{\sigma}f_{\sigma}(\zeta_{3},\ldots,\zeta_{n})=0.
\end{align}
Reducing denominators in 
this equation, the condition for the existence of planar zeros is recast into a homogeneous polynomial
of degree ${1\over 2}(n-2)(n-3)$ in the stereographic variables. 

In the case of graviton scattering, the analysis carried out in \cite{ours_pz} showed that 
the planar, five-graviton amplitude automatically vanishes without imposing any further condition on the 
kinematics of the outgoing particles. 

The aim of this paper is to further investigate these issues, focusing on the
 conditions under which the projective structure of the planar zeros
is preserved. We will see that the equation determining them  
is invariant under a simultaneous rescaling of the outgoing stereographic coordinates
for theories with gauge invariance, even
when matter scalar fields are introduced. The resulting projective
curves are of the same type as the ones found for gluon scattering in Ref. \cite{ours_pz}.
Pure scalar theories, on the other hand, give rise to equations for the
existence of planar zeros which are not homogeneous in the stereographic coordinates. 

In the case of graviton scattering, we trace the vanishing of the planar five-graviton amplitude found in \cite{ours_pz}
to the fact that the amplitude becomes effectively three-dimensional in this limit. According to a general result of
\cite{huang_johansson}, odd-multiplicity three-dimensional gravitation amplitudes vanish due to helicity 
non-conservation. This conclusion is confirmed by a computation of the planar limit of the helicity-preserving six-graviton amplitude, which gives a nonzero result. 
We revisit the amplitudes for the scattering of scalars with graviton emission to find
that they also vanish identically in the planar limit, similarly to what happens with 
the five-graviton amplitude computed in \cite{ours_pz}.  

We also study the corrections to planar zeros associated with the ultraviolet completions of gauge theories and gravity
provided by string theory.
For Yang-Mills scattering, we compute the five gauge bosons disk amplitude using the methods developed in 
Ref. \cite{MSS}. Expanding the generalized Euler integrals in powers of the inverse string tension, we find that 
the planar zero condition found in the field theory limit gets corrected by equations which fail to preserve 
the projective structure found in \cite{ours_pz}. 
In the context of gravity amplitudes, 
we compute the five-graviton closed string amplitude on the sphere and its $\alpha'$ expansion 
using the single-valued projection \cite{stieberger_taylor,svp,svp2}. Unlike its field theory limit, the planar 
graviton string amplitude is generically nonzero, thus avoiding the consequences of the theorem proved in \cite{huang_johansson}. 

The plan of the paper is as follows. 
In Section \ref{sec:phi3} we analyze planar zeros in a cubic non-gauge massless scalar theory transforming 
as bi-adjoints under a global symmetry group. We particularize our analysis to the case of a  
$\phi^{3}$ theory, where we study the structure of the curves determining the planar zeros. 
We dedicate Section \ref{sec:scalar_QCD} to the study of planar zeros in a theory of scalars 
coupled to a gauge field. Having completed our presentation of the properties of planar zeros in gauge field theories, we 
proceed in Section \ref{sec:string} to the computation of the $\alpha'$ corrections to the five-gluon
amplitude in the planar limit.  

In Section \ref{sec:grav_scalars} we turn our attention to the planar zeros of gravitational scattering amplitudes,
considering the collision of two scalars, both distinguishable and indistinguishable, with emission of a graviton. 
Here we also analyze the helicity-preserving six-graviton 
amplitude, which turns out to be nonzero in the planar limit.  Section \ref{sec:closed_strings}
is devoted to the study of the $\alpha'$ corrections to the planar five graviton tree level amplitude. 
Finally, in Section \ref{sec:soft} we discuss how the projective properties
of planar zeros in theories with gauge invariance emerge from the structure of the amplitude in the soft limit.
Our conclusions are summarized in Section \ref{sec:closing}.
To avoid cluttering the main text with cumbersome expressions, some long equations have been deferred to the Appendix.

\section{Pure scalar theories}
\label{sec:phi3}

We begin with the analysis of a pure scalar theory transforming in the biadjoint representation of a generic
global symmetry group $G\times \overline{G}$, with action
\begin{align}
S=\int d^{4}x\left({1\over 2}\partial_{\mu}\Phi^{a\overline{a}}\partial^{\mu}\Phi^{a\overline{a}}-
{\lambda\over 3!}f^{abc}\overline{f}^{\overline{a}\overline{b}\overline{c}}\Phi^{a\overline{a}}\Phi^{b\overline{b}}
\Phi^{c\overline{c}}\right),
\end{align}
and study the tree-level, five point amplitude. A very economic way of obtaining this amplitude 
is by using the so-called zeroth-copy prescription \cite{zero_copy}, consisting in replacing kinematic 
numerators in the pure-gauge theory amplitude with a second copy of the color factors
\begin{align}
\mathcal{A}_{n,\rm gauge}=g_{\rm YM}^{n-2}\sum_{i\in\Gamma}{c_{i}n_{i}\over \prod_{\alpha} s_{i,\alpha}}
\hspace*{1cm} \Longrightarrow \hspace*{1cm}
\mathcal{A}_{n,\rm scalar}=\lambda^{n-2}\sum_{i\in\Gamma}{c_{i}\overline{c}_{i}\over \prod_{\alpha} s_{i,\alpha}},
\end{align}
where $c_{i}$, $\overline{c}_{i}$ are the color factors of $G$ and $\overline{G}$ respectively.
With this, we find
\begin{align}
\mathcal{A}_{5,\rm scalars}&=i\lambda^{3}\left({c_{1}\overline{c}_{1}\over s_{12}s_{45}}+{c_{2}\overline{c}_{2}\over s_{23}s_{15}}
+{c_{3}\overline{c}_{3}\over s_{34}s_{12}}
+{c_{4}\overline{c}_{4}\over s_{45}s_{23}}+{c_{5}\overline{c}_{5}\over s_{15}s_{34}}+{c_{6}\overline{c}_{6}\over s_{14}s_{25}}
+{c_{7}\overline{c}_{7}\over s_{13}s_{25}}+{c_{8}\overline{c}_{8}\over s_{24}s_{13}} \right. \nonumber \\[0.2cm]
&+ \left.{c_{9}\overline{c}_{9}\over s_{35}s_{24}}+{c_{10}\overline{c}_{10}\over s_{14}s_{35}}
+{c_{11}\overline{c}_{11}\over s_{15}s_{24}}
+{c_{12}\overline{c}_{12}\over s_{12}s_{35}}+{c_{13}\overline{c}_{13}\over s_{23}s_{14}}
+{c_{14}\overline{c}_{14}\over s_{25}s_{34}}
+{c_{15}\overline{c}_{15}\over s_{13}s_{45}}\right),
\label{eq:amplitude_general_biadjoint}
\end{align}
where we have defined the kinematic invariants 
\begin{align}
s_{ij}=(p_{i}+p_{j})^{2}=2p_{i}\cdot p_{j} \hspace*{0.5cm} \mbox{where} \hspace*{0.5cm} i<j,
\end{align}
and the color factors are given by 
\begin{align}
c_{1} &= f^{a_{1}a_{2}b} f^{ba_{3}c} f^{ca_{4}a_{5}}, \hspace*{1cm} c_{2}=f^{a_{1}a_{5}b} f^{ba_{4}c} f^{ca_{3}a_{2}}, \hspace*{1cm}
c_{3} = f^{a_{3}a_{4}b} f^{ba_{5}c} f^{ca_{1}a_{2}}, 
\nonumber \\[0.2cm]
 \hspace*{1cm} c_{4}&=f^{a_{4}a_{5}b} f^{ba_{1}c} f^{ca_{2}a_{3}},
\hspace*{1cm} 
c_{5} = f^{a_{5}a_{1}b} f^{ba_{2}c} f^{ca_{3}a_{4}}, \hspace*{1cm} c_{6}=f^{a_{1}a_{4}b} f^{ba_{3}c} f^{ca_{5}a_{2}},
\nonumber \\[0.2cm]
c_{7} &= f^{a_{1}a_{3}b} f^{ba_{4}c} f^{ca_{5}a_{2}}, \hspace*{1cm} c_{8}=f^{a_{1}a_{3}b} f^{ba_{5}c} f^{ca_{4}a_{2}}, \hspace*{1cm} 
c_{9} = f^{a_{3}a_{5}b} f^{ba_{1}c} f^{ca_{2}a_{4}}, 
\label{eq:color_factors} \\[0.2cm]
c_{10}&=f^{a_{4}a_{1}b} f^{ba_{2}c} f^{ca_{3}a_{5}},
\hspace*{0.85cm}
c_{11} = f^{a_{1}a_{5}b} f^{ba_{3}c} f^{ca_{4}a_{2}}, \hspace*{0.85cm} c_{12}=f^{a_3a_5b} f^{ba_4c} f^{ca_1a_2},
\nonumber \\[0.2cm]
c_{13} &= f^{a_{1}a_{4}b} f^{ba_{5}c} f^{ca_{3}a_{2}}, \hspace*{0.85cm} c_{14}=f^{a_{5}a_{2}b} f^{ba_1c} f^{ca_{3}a_{4}},
\hspace*{0.85cm}
c_{15} = f^{a_{1}a_{3}b} f^{ba_{2}c} f^{ca_{4}a_{5}}.
\nonumber
\end{align}

As in \cite{ours_pz}, we work in a center-of-mass reference frame in which the incoming particles have momenta
given by \eqref{eq:momentum_parametrization_in},
whereas the momenta of the three outgoing particles are parametrized using the stereographic coordinates as
given in Eq. \eqref{eq:momentum_parametrization_out}. In our convention
all momenta enter the diagram. We consider planar scattering processes taking place on the plane $y=0$, i.e.
$\zeta_{a}=\overline{\zeta}_{a}$ for $a=3,4,5$. In the case of the five-point amplitude, imposing energy-momentum conservation completely determines the energies of
the outgoing particles: 
\begin{align}
\omega_{3}&={\sqrt{s}\over 2}{(1+\zeta_{3}^{2})(1+\zeta_{4}\zeta_{5})\over (\zeta_{3}-\zeta_{4})(\zeta_{3}-\zeta_{5})}, 
\nonumber \\[0.2cm]
\omega_{4}&={\sqrt{s}\over 2}{(1+\zeta_{4}^{2})(1+\zeta_{3}\zeta_{5})\over (\zeta_{4}-\zeta_{3})(\zeta_{4}-\zeta_{5})},
\label{eq:energies_dehomog}\\[0.2cm]
\omega_{5}&={\sqrt{s}\over 2}{(1+\zeta_{5}^{2})(1+\zeta_{3}\zeta_{4})\over (\zeta_{5}-\zeta_{3})(\zeta_{5}-\zeta_{4})}.
\nonumber
\end{align}
Writing the kinematic invariants in \eqref{eq:amplitude_general_biadjoint} using our parametrization of the momenta, we arrive at the 
following equation for the {\rm planar} five-point scalar amplitude
\begin{align}
\mathcal{A}_{5,\rm scalar}\Bigg|_{\rm planar}=\left({i\lambda^{3}\over s^{2}}\right){P_{10}(\zeta_{3},\zeta_{4},\zeta_{5})\over \zeta_{3}^{2}\zeta_{4}^{2}\zeta_{5}^{2}
(1+\zeta_{3}\zeta_{4})(1+\zeta_{3}\zeta_{5})(1+\zeta_{4}\zeta_{5})},
\label{eq:scalar_amplitude_gen}
\end{align}
where $P_{10}(\zeta_{3},\zeta_{4},\zeta_{5})$ is a degree-ten polynomial whose coefficients depend on 
the color factors. The explicit expression for this polynomial can be found in Eq.~(\ref{sec:appendixP10}). 
The amplitude has collinear singularities at $\zeta_{a}\rightarrow 0,\infty$ together with soft poles at
$\zeta_{a}\zeta_{b}\rightarrow -1$ (with $a<b$) where the energy of one of the outgoing particles tend to 
zero. In addition, the collinear limits $\zeta_{a}\rightarrow \zeta_{b}$ lead to a 
divergence of the energies of the outgoing particles. 

Planar zeros are thus determined by the equation
\begin{eqnarray}
P_{10}(\zeta_{3},\zeta_{4},\zeta_{5})=0.
\end{eqnarray}
Inspecting Eq. \eqref{eq:P10_general_biadjoing}, we find that, unlike the case of pure gauge theories
studied in \cite{ours_pz}, this equation is not homogeneous in the stereographic coordinates, since 
it contains monomials of both degree 10 and 8. Thus, unlike the case of pure gauge theories studied in
\cite{ours_pz}, planar zeros are no longer determined by a projective curve\footnote{Still, looking at
\eqref{eq:P10_general_biadjoing}, we see that the equation $P_{10}(\zeta_{3},\zeta_{4},\zeta_{5})=0$ is
homogeneous in the color factors. As a consequence, after a proper normalization of the group theory
generators, planar zeros are determined by an equation with integer coefficients.}.

\begin{figure}[t]
\centerline{\includegraphics[scale=0.45]{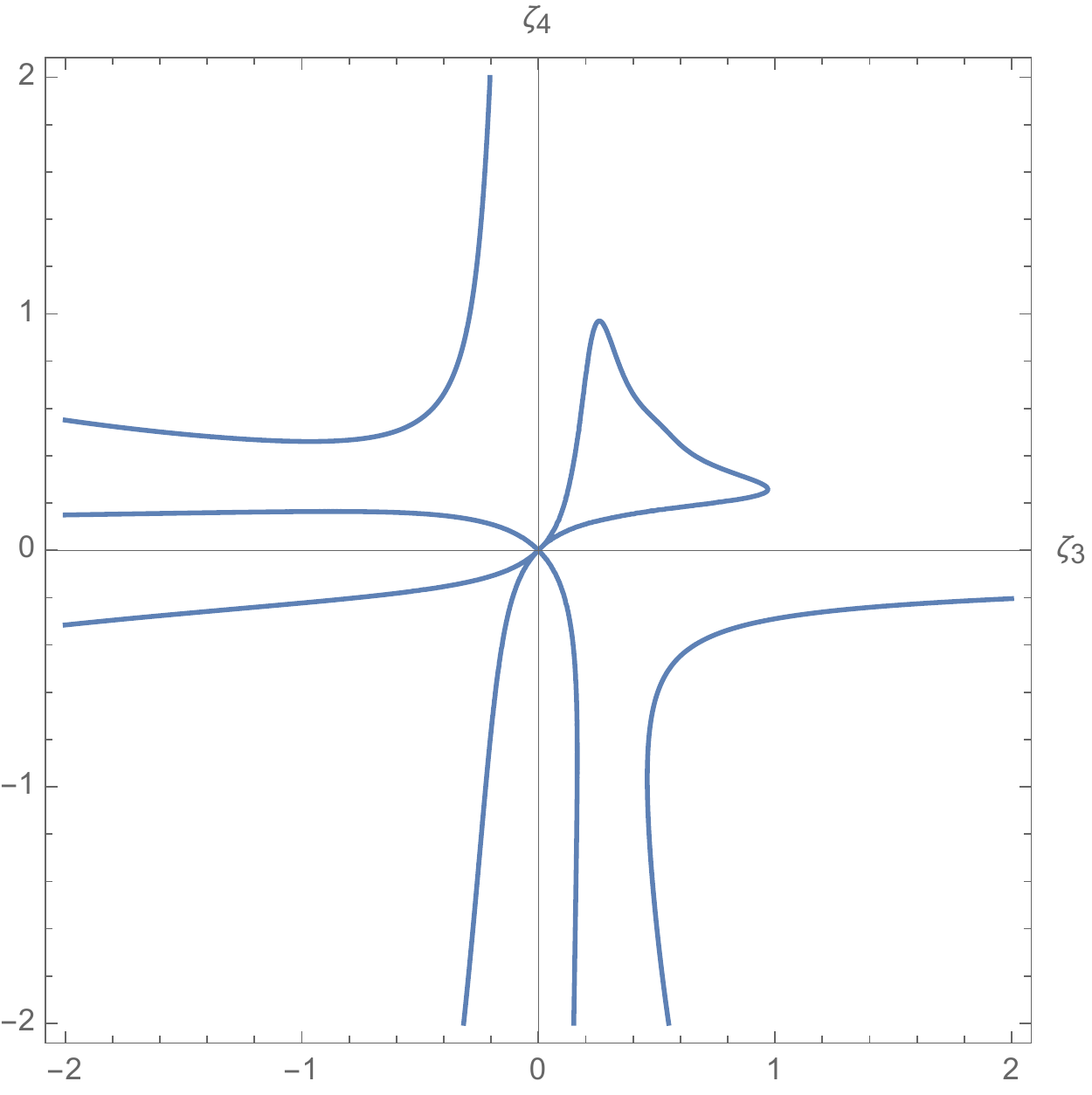}\hspace*{1cm}\includegraphics[scale=0.45]{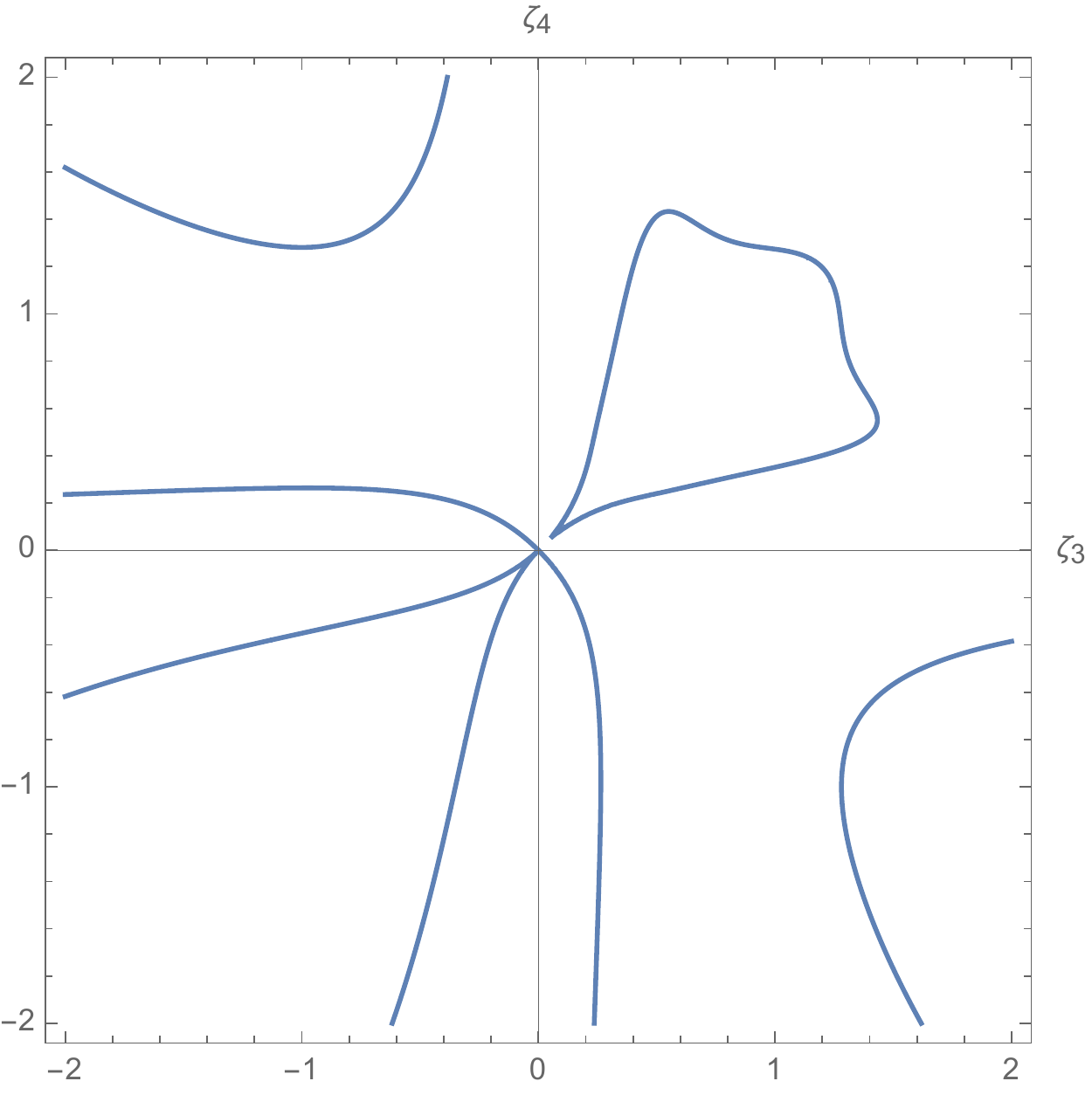}}
\vspace*{1cm}
\centerline{\includegraphics[scale=0.45]{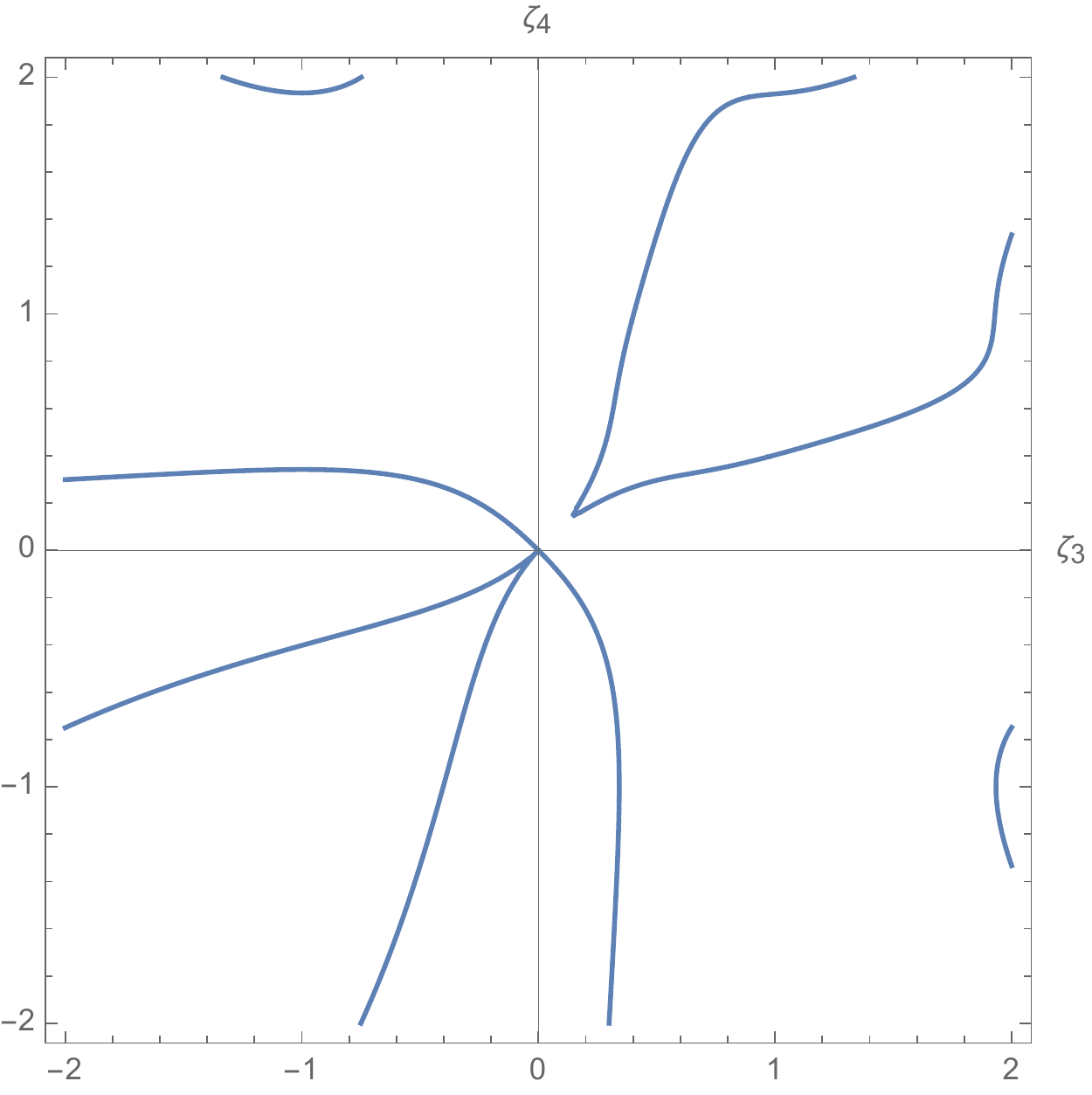}\hspace*{1cm}\includegraphics[scale=0.45]{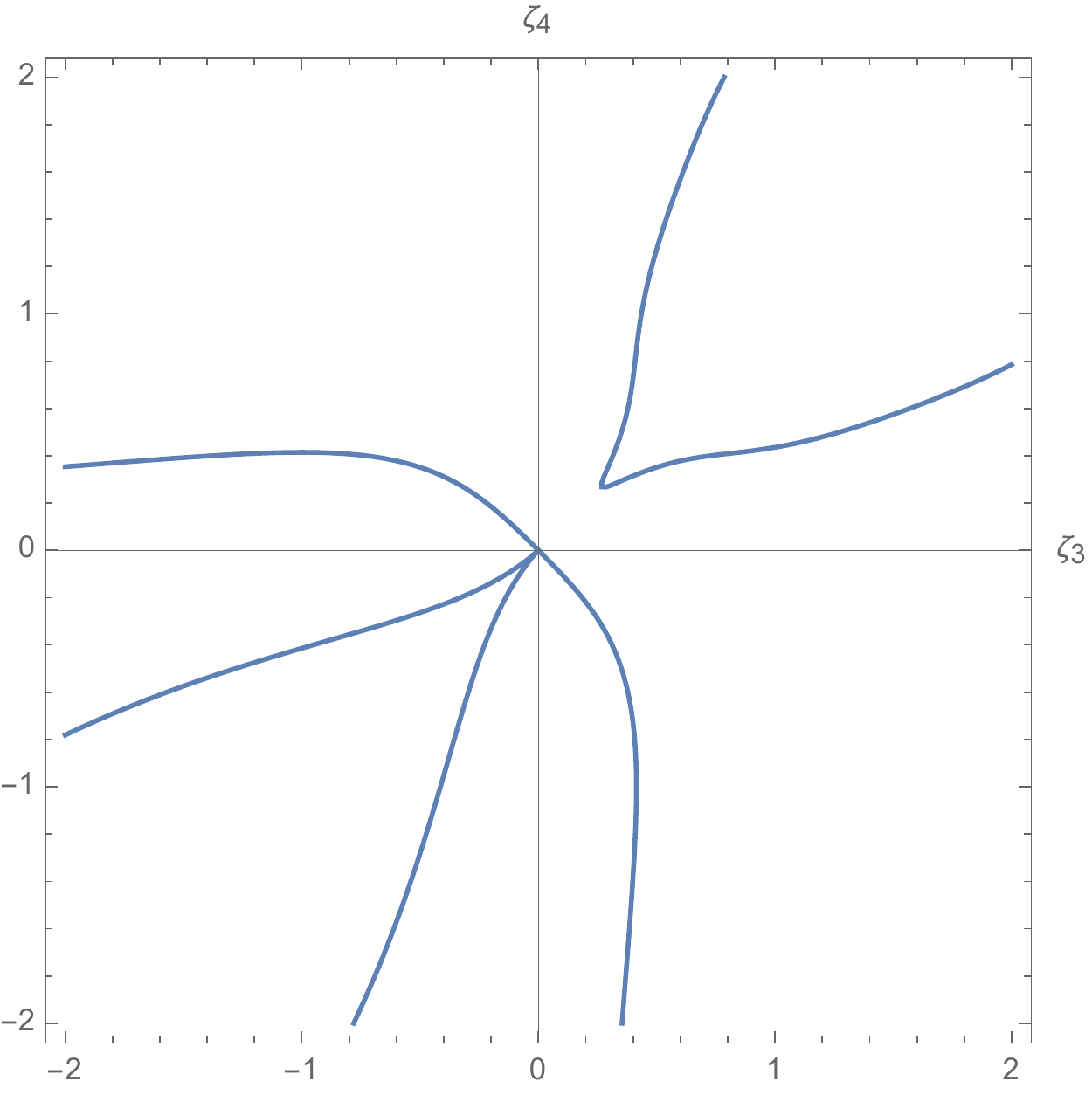}}
\caption[]{Sections of the loci of planar zeros $P_{10}^{\phi^{3}}(\zeta_{3},\zeta_{4},\zeta_{5})=0$ 
for $\zeta_{5}={1\over 4}, {1\over 2}, {3\over 4}, 1$.}
\label{fig:sections}
\end{figure}

To study the corresponding geometric loci of planar zeros, we
focus on the five-point amplitude for $\phi^{3}$ theory, which can be retrieved from Eq. 
\eqref{eq:amplitude_general_biadjoint} by setting all color factors equal to one, $c_{i}\overline{c}_{i}
\rightarrow 1$. In this case, the expression for $P_{10}(\zeta_{3},\zeta_{4},\zeta_{5})$ in 
\eqref{eq:scalar_amplitude_gen} somewhat simplifies to 
\begin{align}
P_{10}^{\phi^{3}}(\zeta_{3},\zeta_{4},\zeta_{5})&=
\zeta_{3}^{6}\zeta_{4}^{3} \zeta_{5}-\zeta_{3}^{6} \zeta_{4}^{2}\zeta_{5}^{2}+\zeta_{3}^{6}\zeta_{4}\zeta_{5}^{3}
-\zeta_{3}^{5}\zeta_{4}^{4}\zeta_{5}-\zeta_{3}^{5}\zeta_{4}^{3}\zeta_{5}^{2}+\zeta_{3}^{5}\zeta_{4}^{3}
\nonumber \\[0.2cm]
&-\zeta_{3}^{5}\zeta_{4}^{2}\zeta_{5}^{3}-\zeta_{3}^{5}\zeta_{4}^{2}\zeta_{5}-\zeta_{3}^{5}\zeta_{4}\zeta_{5}^{4}
-\zeta_{3}^{5}\zeta_{4}\zeta_{5}^{2}+\zeta_{3}^{5}\zeta_{5}^{3}-\zeta_{3}^{4}\zeta_{4}^{5}\zeta_{5}
   \nonumber \\[0.2cm]
&+4\zeta_{3}^{4}\zeta_{4}^{4}\zeta_{5}^{2}-\zeta_{3}^{4}\zeta_{4}^{4}-\zeta_{3}^{4}\zeta_{4}^{3}\zeta_{5}^{3}
-\zeta_{3}^{4}\zeta_{4}^{3}\zeta_{5}+4\zeta_{3}^{4}\zeta_{4}^{2}\zeta_{5}^{4}
+4\zeta_{3}^{4}\zeta_{4}^{2}\zeta_{5}^{2}
   \nonumber \\[0.2cm]
&-\zeta_{3}^{4}\zeta_{4}\zeta_{5}^{5}-\zeta_{3}^{4}\zeta_{4}\zeta_{5}^{3}-\zeta_{3}^{4}\zeta_{5}^{4}
+\zeta_{3}^{3}\zeta_{4}^{6}\zeta_{5}-\zeta_{3}^{3}\zeta_{4}^{5}\zeta_{5}^{2}+\zeta_{3}^{3}\zeta_{4}^{5}\nonumber \\[0.2cm]
&-\zeta_{3}^{3} \zeta_{4}^{4}
   \zeta_{5}^{3}-\zeta_{3}^{3}\zeta_{4}^{4}\zeta_{5}-\zeta_{3}^{3}\zeta_{4}^{3}\zeta_{5}^{4}
   -\zeta_{3}^{3}\zeta_{4}^{3}\zeta_{5}^{2}-\zeta_{3}^{3}\zeta_{4}^{2}\zeta_{5}^{5}-\zeta_{3}^{3}\zeta_{4}^{2}\zeta_{5}^{3} 
\label{eq:P10_phi3}  \\[0.2cm]
&+\zeta_{3}^{3}\zeta_{4}\zeta_{5}^{6}-\zeta_{3}^{3}\zeta_{4}\zeta_{5}^{4}+\zeta_{3}^{3}\zeta_{5}^{5}
-\zeta_{3}^{2}\zeta_{4}^{6}\zeta_{5}^{2}-\zeta_{3}^{2}\zeta_{4}^{5}\zeta_{5}^{3}-\zeta_{3}^{2}\zeta_{4}^{5}\zeta_{5}\nonumber \\[0.2cm]
&+4\zeta_{3}^{2}\zeta_{4}^{4}\zeta_{5}^{4}+4\zeta_{3}^{2}\zeta_{4}^{4}\zeta_{5}^{2}
-\zeta_{3}^{2}\zeta_{4}^{3}\zeta_{5}^{5}-\zeta_{3}^{2}\zeta_{4}^{3}\zeta_{5}^{3}-\zeta_{3}^{2}\zeta_{4}^{2}\zeta_{5}^{6}
   \nonumber \\[0.2cm]
&+4\zeta_{3}^{2}\zeta_{4}^{2}\zeta_{5}^{4}-\zeta_{3}^{2}\zeta_{4}\zeta_{5}^{5}+\zeta_{3}\zeta_{4}^{6}\zeta_{5}^{3}-\zeta_{3}
   \zeta_{4}^{5}\zeta_{5}^{4}-\zeta_{3}\zeta_{4}^{5}\zeta_{5}^{2}-\zeta_{3}\zeta_{4}^{4}\zeta_{5}^{5}
  \nonumber \\[0.2cm]
&-\zeta_{3}\zeta_{4}^{4}\zeta_{5}^{3}+\zeta_{3}\zeta_{4}^{3}\zeta_{5}^{6}-\zeta_{3}\zeta_{4}^{3}\zeta_{5}^{4}
-\zeta_{3}\zeta_{4}^{2}\zeta_{5}^{5}+\zeta_{4}^{5}\zeta_{5}^{3}-\zeta_{4}^{4}\zeta_{5}^{4}+\zeta_{4}^{3}\zeta_{5}^{5}. 
\nonumber
\end{align}
Notice that the polynomial is symmetric in all its entries, as 
expected by Bose symmetry. This implies that when studying the curves, we can consider sections of 
constant $\zeta_{5}$ without loss of generality. In Fig. \ref{fig:sections} we have plotted the 
sections for $\zeta_{5}={1\over 4}$, ${1\over 2}$, ${3\over 4}$, and $1$. The curves are more complicated
than in the pure gauge case and include singular points.

\section{Planar zeros in scalar QCD}
\label{sec:scalar_QCD}

The results of the previous section show that the projective character of planar zeros in the five-point
function gluon amplitude disappears when considering pure scalar theories, even in the presence of global 
symmetries. This is indeed due to the absence of derivative couplings, which render
the numerators appearing in the scalar amplitude \eqref{eq:amplitude_general_biadjoint} trivial.
It is therefore tempting to conclude that, despite the similarities in the topologies contributing to both
the gauge and scalar amplitudes, the projective nature of the equation determining the planar zeros is
a consequence of gauge invariance.

\subsection{Distinguishable scalars}
\label{sec:disct_scalars_gauge}

To further explore this possibility, we 
study now the presence of planar zeros in scalar QCD (sQCD), in particular in the scattering of two distinct scalars
with emission of a gluon in the final state. This process has been studied in Ref. \cite{sv_sc_vm2}. 
We label momenta and color quantum numbers 
according to
\begin{align}
\Phi(p_{1},j)+\Phi'(p_{2},n)\longrightarrow \Phi(p_{3},i)+\Phi'(p_{4},m)+g(p_{5},a,\epsilon_{+}),
\end{align}
where $\epsilon_{+}(p_{5})$ indicates the polarization vector of the gluon. We slightly modify the
conventions of Ref. \cite{sv_sc_vm2}, 
and consider all momenta as incoming. The amplitude takes the form
\begin{align}
\mathcal{A}=g^{3}\left({C_{1}n_{1}\over s_{24}s_{35}}+{C_{2}n_{2}\over s_{24}s_{15}}+{C_{3}n_{3}\over s_{24}}
+{C_{4}n_{4}\over s_{13}s_{45}}+{C_{5}n_{5}\over s_{13}s_{25}}+{C_{6}n_{6}\over s_{13}}+{C_{7}n_{7}\over s_{13}s_{24}}\right),
\label{eq:sQCD_amp}
\end{align}
where now there are
seven different color factors 
\begin{align}
C_{1}= T^{a}_{ik}T^{b}_{kj}\overline{T}^{b}_{mn}, \hspace*{0.8cm}  
C_{2}= T^{b}_{ik}T^{a}_{kj}\overline{T}^{b}_{mn}, \hspace*{0.8cm} &
C_{3}= T^{a}_{ik}T^{b}_{kj}\overline{T}^{b}_{mn}+T^{b}_{ik}T^{a}_{kj}\overline{T}^{b}_{mn}, \nonumber \\[0.2cm]
C_{4}= T^{b}_{ij}\overline{T}^{a}_{mk}\overline{T}^{b}_{kn}, \hspace*{0.7cm}
C_{5}= T^{b}_{ij}\overline{T}^{b}_{mk}\overline{T}^{a}_{kn}, \hspace*{0.8cm} &
C_{6}= T^{b}_{ij}\overline{T}^{a}_{mk}\overline{T}^{b}_{kn}+T^{b}_{ij}\overline{T}^{b}_{mk}\overline{T}^{a}_{kn},  \\[0.2cm]
C_{7}= if^{abc}T^{b}_{ij}\overline{T}^{c}_{mn}. \hspace*{4.4cm} &\nonumber
\end{align}
We allow for the possibility of the two scalars transforming in different representations of the
gauge group. The color factors satisfy four Jacobi identities
\begin{align}
C_{1}-C_{2}+C_{7}=0,  & \hspace*{1cm}
C_{1}+C_{2}-C_{3}=0, \label{eq:jacobi_id_dist_scalars_Cs} \\[0.2cm]
C_{4}-C_{5}-C_{7}=0, &\hspace*{1cm}
C_{4}+C_{5}-C_{6}=0. \nonumber
\end{align}
These relations can be used to express the five-point amplitudes
in terms of only three independent color factors, that we take $C_{1}$, $C_{2}$, and $C_{4}$. Namely,
\begin{align}
C_{3}= C_{1}+C_{2},\hspace*{1.2cm} &\hspace*{1cm}
C_{5}= C_{1}-C_{2}+C_{4}, \nonumber \\[0.2cm]
C_{6}= C_{1}-C_{2}+2C_{4}, &\hspace*{1cm}
C_{7}= -C_{1}+C_{2}. 
\end{align}

Again, we work in the center-of-mass reference frame and use \eqref{eq:momentum_parametrization_out} to 
express outgoing momenta in terms of the stereographic coordinates. In the planar limit 
$\zeta_{a}=\overline{\zeta}_{a}$ we take the gluon polarization vector
to be 
\begin{eqnarray}
\epsilon_{\pm}=\pm {1\over\sqrt{2}}\left(0,{\zeta_{5}^{2}-1\over 1+\zeta_{5}^{2}},\mp i,-{2\zeta_{5}\over 1+\zeta_{5}^{2}}\right),
\label{eq:gluon_pol}
\end{eqnarray} 
which indeed satisfies $p_{5}\cdot\epsilon_{\pm}(p_{5})=0$. 
In the following, we specialize our analysis to a positive helicity gluon, 
$\epsilon\equiv \epsilon_{+}$. With this, the numerators in the 
amplitude \eqref{eq:sQCD_amp} take the following form in the planar limit
\begin{align}
n_{1}&= -2i[(p_{1}-p_{3}-p_{5})\cdot (p_{2}-p_{4})][p_{3}\cdot\epsilon_{+}(p_{5})] \nonumber \\[0.2cm]
&={i\sqrt{2}s^{3\over 2}(1+\zeta_{3}\zeta_{5})(1+\zeta_{4}\zeta_{5})\over (1+\zeta_{5}^{2})
(\zeta_{3}-\zeta_{4})^{2}(\zeta_{4}-\zeta_{5})}(-1+\zeta_{3}\zeta_{4}-\zeta_{4}^{2}-2\zeta_{3}\zeta_{5}
+\zeta_{4}\zeta_{5}), \nonumber  \\[0.2cm]
n_{2}&= 2i[(p_{1}-p_{3}+p_{5})\cdot (p_{2}-p_{4})][p_{1}\cdot \epsilon_{+}(p_{5})] \nonumber \\[0.2cm]
&= {i\sqrt{2}s^{3\over 2}\zeta_{5}(1+\zeta_{4}\zeta_{5})\over (1+\zeta_{5}^{2})
(\zeta_{3}-\zeta_{4})(\zeta_{3}-\zeta_{5})(\zeta_{4}-\zeta_{5})}(-\zeta_{3}+\zeta_{4}+\zeta_{3}^{2}\zeta_{4}
-2\zeta_{3}\zeta_{5}+\zeta_{3}\zeta_{4}\zeta_{5}),
\nonumber \\[0.2cm]
n_{3}&= -i(p_{2}-p_{4})\cdot\epsilon_{+}(p_{5}) \nonumber \\[0.2cm] 
&= {i\sqrt{s}\over \sqrt{2}(1+\zeta_{5}^{2})(\zeta_{3}-\zeta_{4})}(1+2\zeta_{3}\zeta_{5}
+\zeta_{3}\zeta_{4}\zeta_{5}^{2}), \nonumber \\[0.2cm]
n_{4}&= -2i[(p_{2}-p_{4}-p_{5})\cdot (p_{1}-p_{3})][p_{4}\cdot \epsilon_{+}(p_{5})] \\[0.2cm]
&= -{i\sqrt{2}s^{3\over 2}(1+\zeta_{3}\zeta_{5})(1+\zeta_{4}\zeta_{5})\over (1+\zeta_{5}^{2})
(\zeta_{3}-\zeta_{4})^{2}(\zeta_{3}-\zeta_{5})}(2\zeta_{3}^{2}-\zeta_{3}\zeta_{4}-\zeta_{3}\zeta_{5}
+\zeta_{4}\zeta_{5}+\zeta_{3}^{2}\zeta_{4}\zeta_{5}),
\nonumber \\[0.2cm]
n_{5}&= 2i[(p_{2}-p_{4}+p_{5})\cdot (p_{1}-p_{3})][p_{2}\cdot \epsilon_{+}(p_{5})] \nonumber \\[0.2cm]
&= -{i\sqrt{2}s^{3\over 2}\zeta_{5}(1+\zeta_{3}\zeta_{5})\over (1+\zeta_{5}^{2})(\zeta_{3}-\zeta_{4})
(\zeta_{3}-\zeta_{5})(\zeta_{4}-\zeta_{5})}(-2\zeta_{3}+\zeta_{4}+\zeta_{5}-\zeta_{3}\zeta_{4}\zeta_{5}
+\zeta_{4}^{2}\zeta_{5}), 
\nonumber \\[0.2cm]
n_{6}&= -i(p_{1}-p_{3})\cdot \epsilon_{+}(p_{5}) \nonumber \\[0.2cm]
&= -{i\sqrt{s}\over \sqrt{2}(1+\zeta_{5}^{2})(\zeta_{3}-\zeta_{4})}
(1+2\zeta_{3}\zeta_{5}+\zeta_{3}\zeta_{4}\zeta_{5}^{2}), 
\nonumber \\[0.2cm]
n_{7}&= -i\Big\{[(p_{2}-p_{4})\cdot (p_{1}+p_{3}-p_{5})][(p_{1}-p_{3})\cdot \epsilon_{+}(p_{5})] \nonumber \\[0.2cm]
& -\,\,[(p_{1}-p_{3})\cdot (p_{2}-p_{4})][(p_{3}+p_{1}-p_{4}-p_{2})\cdot \epsilon_{+}(p_{5})] \nonumber \\[0.2cm]
& -\,\,[(p_{1}-p_{3})\cdot (p_{4}+p_{2}-p_{5})][(p_{2}-p_{4})\cdot \epsilon_{+}(p_{5})]\Big\} \nonumber
\\[0.2cm]
&={is^{3\over 2}\over \sqrt{2}(1+\zeta_{5}^{2})(\zeta_{3}-\zeta_{4})^{2}(\zeta_{3}-\zeta_{5})
(\zeta_{4}-\zeta_{5})}\Big(
-\zeta_{3}^{3}\zeta_{4}^{3}\zeta_{5}^{3}-4\zeta_{3}^{3}\zeta_{4}^{2}\zeta_{5}^2
-4\zeta_{3}^{3}\zeta_{4}\zeta_{5} \nonumber \\[0.2cm]
&+\zeta_{3}^{2}\zeta_{4}^{3}
\zeta_{5}^{4}+\zeta_{3}^{2}\zeta_{4}^{3}\zeta_{5}^{2}
+8\zeta_{3}^{2}\zeta_{4}^{2}\zeta_{5}^{3}
-\zeta_{3}^{2}\zeta_{4}^{2}\zeta_{5}-4\zeta_{3}^{2}\zeta_{4}\zeta_{5}^4
+8\zeta_{3}^{2}\zeta_{4}\zeta_{5}^{2}-4\zeta_{3}^{2}\zeta_{4} \nonumber \\[0.2cm]
&-4\zeta_{3}^{2}\zeta_{5}^{3}
-4\zeta_{3}\zeta_{4}^{3}\zeta_{5}^{3}+2\zeta_{3}\zeta_{4}^{3}\zeta_{5} 
+3\zeta_{3}\zeta_{4}^{2}\zeta_{5}^{4}+2\zeta_{3}\zeta_{4}^{2}\zeta_{5}^{2}+
3\zeta_{3}\zeta_{4}^{2}-\zeta_{3}\zeta_{4}\zeta_{5}^{3} \nonumber \\[0.2cm]
&+8\zeta_{3}\zeta_{4}\zeta_{5}
-4\zeta_{3}\zeta_{5}^{2}-4\zeta_{4}^{3}\zeta_{5}^{2} 
+2\zeta_{4}^{2}\zeta_{5}^{3}-4\zeta_{4}^{2}\zeta_{5}+\zeta_{4}\zeta_{5}^{2}+\zeta_{4}-\zeta_{5}\Big).
\nonumber
\end{align}
As a nontrivial test of the previous equations, it can be checked that the amplitude satisfies the gauge Ward identity. Combining the numerators with the expressions for the kinematic invariants we arrive at the 
following form of the tree-level amplitude of distinct scalars
with gluon emission in the limit of planar scattering:
\begin{align}
\mathcal{A}_{5}\Bigg|_{\rm planar}
={i\sqrt{2}g^{3}(2\zeta_{3}-\zeta_{4}) \over \sqrt{s}\zeta_{4}\zeta_{5}(1+\zeta_{3}\zeta_{4})}
\Big[(C_{1}-C_{2}+C_{4})\zeta_{3}\zeta_{4}-(C_{1}-C_{2})\zeta_{3}\zeta_{5}
+(C_{2}-C_{4})\zeta_{4}\zeta_{5}-C_{2}\zeta_{5}^{2}\Big].
\end{align}

Similarly to what happens for pure gauge theories \cite{ours_pz}, 
planar zeros are determined by a homogeneous cubic polynomial
\begin{align}
(2\zeta_{3}-\zeta_{4})\Big[(C_{1}-C_{2}+C_{4})\zeta_{3}\zeta_{4}-(C_{1}-C_{2})\zeta_{3}\zeta_{5}
+(C_{2}-C_{4})\zeta_{4}\zeta_{5}-C_{2}\zeta_{5}^{2}\Big]=0.
\label{eq:full_cubic_sQCD}
\end{align}
An important difference with the pure gauge theory case is, however, 
that the polynomial factorizes. One of the factors, the trivial 
branch, is linear and independent of the color factors of the interacting particles,
\begin{align}
2\zeta_{3}-\zeta_{4}=0.
\label{eq:trivial_condition_sQCD_1_full}
\end{align}
It is also independent of the direction of flight of the emitted gluon. 
The second, non-trivial branch is a quadratic equation
\begin{align}
(C_{1}-C_{2}+C_{4})\zeta_{3}\zeta_{4}-(C_{1}-C_{2})\zeta_{3}\zeta_{5}
+(C_{2}-C_{4})\zeta_{4}\zeta_{5}-C_{2}\zeta_{5}^{2}=0,
\label{eq:condition_sQCD_2_full_zetas}
\end{align}
whose coefficients depend on the three independent color factors. 

Being also homogeneous in the color factors, the 
polynomial \eqref{eq:full_cubic_sQCD} defines an integer curve in the projective plane defined by 
the coordinates $(\zeta_{3},\zeta_{4},\zeta_{5})$. It seems natural now to single out 
the direction of flight of the emitted gluon and study this curve in the patch centered around 
the point $(0,0,1)$ using the coordinates
\begin{align}
(\zeta_{3},\zeta_{4},\zeta_{5})=\lambda(U,V,1).
\label{eq:projective_coordinates1}
\end{align}
Now, the trivial branch of planar zeros is determined by the straight line
\begin{align}
2U-V= 0,
\label{eq:condition_sQCD_1}
\end{align}
whereas the non-trivial quadratic curve takes the form
\begin{align}
(C_{1}-C_{2}+C_{4})UV-(C_{1}-C_{2})U
+(C_{2}-C_{4})V-C_{2}=0.
\label{eq:condition_sQCD_2}
\end{align}

The quadratic curve \eqref{eq:condition_sQCD_2} can be easily classified for a generic gauge group
in terms of the 
three invariants ($\Delta$, $\delta$, $I$) and the semiinvariant $\sigma$ 
(see, for example, \cite{handbook}) defined by 
\begin{align}
\Delta &= {1\over 4}C_{1}C_{4}(C_{1}-C_{2}+C_{4}), \nonumber \\[0.2cm]
\delta&= -{1\over 4}(C_{1}-C_{2}+C_{4})^{2}, \nonumber \\[0.2cm]
I&= 0, \\[0.2cm]
\sigma&= -{1\over 4}(C_{1}-C_{2})^{2}-{1\over 4}(C_{2}-C_{4})^{2}. \nonumber
\end{align}
Since $\delta\leq 0$, no ellipses are possible. It is also impossible to have $\delta=0$ with $\Delta\neq 0$, so 
parabolas are ruled out as well. Thus, the only possible class of curves are 
hyperbolas ($\Delta\neq 0$, $\delta<0$),
intersecting lines ($\Delta=0$, $\delta<0$), or parallel lines ($\Delta=\delta=0$, $\sigma<0$). Notice that this
classification is valid for {\it all} gauge groups and {\it all} representations of the scalar fields. \label{page:general_analysis}

As an illustrative example, we study the case of two scalars with charges $e$ and $e'$ coupled to a photon. 
This correspond to having the U(1) generators
\begin{align}
T^{1}_{ij}=e\delta_{ij}, \hspace*{1cm} \overline{T}^{1}_{mn}=e'\delta_{mn}, 
\end{align}
giving the following values for the color factors
\begin{align}
C_{1}=C_{2}={1\over 2}C_{3}=e^{2}e', \hspace*{1cm}
C_{4}=C_{5}={1\over 2}C_{6}=ee'^{\,2}, \hspace*{1cm} C_{7}=0.
\end{align}
In the patch centered around $U=V=0$, the projective curve determining the planar zeros is given by
\begin{align}
ee'^{2}UV+ee'(e-e')V-e^{2}e'=0.
\label{eq:maxwell_zeros}
\end{align}
Since 
\begin{align}
\Delta={1\over 4}e^{4}e'^{\,5}\neq 0, \hspace*{1cm} \delta=-{1\over 4}e^{2}e'^{4}<0,
\end{align}
the loci of planar zeros are hyperbolas with asymptotes along the coordinates axes and whose center is located at
the point
\begin{align}
(U_{0},V_{0})=\left({e'-e\over e'},0\right).
\end{align}
A particularly simple case arises when we consider that both scalars, though distinct, have the same electric charge, $e=e'$. In this
case the curve is given by $UV=1$. 

\subsection{Indistinguishable scalars}

The previous analysis of the scattering amplitude of distinct scalars coupled to a gauge field in an arbitrary
representation 
illustrates how the derivative couplings required by gauge invariance are enough to restore
the projective nature of planar zeros, that was absent in the pure scalar theories studied in Section 
\ref{sec:phi3}. This is also the case when considering sQCD with a single scalar field in the adjoint
representation of the gauge group. We consider again a five-point amplitude corresponding to 
the process \cite{jsvscvm}
\begin{align}
\Phi(p_{1},a_{1})+\Phi(p_{2},a_{2})\longrightarrow \Phi(p_{3},a_{3})+\Phi(p_{4},a_{4})+g(p_{5},a_{5},\epsilon).
\end{align}
After all quartic couplings are resolved in terms of trivalent vertices, the 
15 topologies contributing to this amplitude are the ones already encountered in both pure Yang-Mills theories
and the scalar theories studied in Section \ref{sec:phi3}. The amplitude takes the form 
\begin{align}
\mathcal{A}_{5}&=g^{3}\left({c_{1}n_{1}\over s_{12}s_{45}}+{c_{2}n_{2}\over s_{23}s_{15}}+{c_{3}n_{3}\over s_{34}s_{12}}
+{c_{4}n_{4}\over s_{45}s_{23}}+{c_{5}n_{5}\over s_{15}s_{34}}+{c_{6}n_{6}\over s_{14}s_{25}}
+{c_{7}n_{7}\over s_{13}s_{25}}+{c_{8}n_{8}\over s_{24}s_{13}} \right. \nonumber \\[0.2cm]
&+ \left.{c_{9}n_{9}\over s_{35}s_{24}}+{c_{10}n_{10}\over s_{14}s_{35}}+{c_{11}n_{11}\over s_{15}s_{24}}
+{c_{12}n_{12}\over s_{12}s_{35}}+{c_{13}n_{13}\over s_{23}s_{14}}+{c_{14}n_{14}\over s_{25}s_{34}}
+{c_{15}n_{15}\over s_{13}s_{45}}\right),
\label{eq:amplitude_ind_scalars_general}
\end{align}
where the color factors are the ones defined in \eqref{eq:color_factors}, while the numerators are given by
\begin{align}
n_{1}&= (p_{4}+p_{5})^{2}[(p_{2}-p_{1})\cdot\epsilon_{+}(p_{5})]
+2(p_{2}-p_{1})\cdot (p_{3}-p_{4}-p_{5})[p_{4}\cdot\epsilon_{+}(p_{5})], \nonumber \\
n_{2}&= -(p_{1}+p_{5})^{2}[(p_{2}-p_{3})\cdot\epsilon_{+}(p_{5})]-2(p_{2}-p_{3})\cdot(-p_{1}+p_{4}-p_{5})
[p_{1}\cdot\epsilon_{+}(p_{5})], \nonumber \\
n_{3}&= -(p_{2}-p_{1})\cdot(p_{3}-p_{4})[(p_{1}+p_{2}-p_{3}-p_{4})\cdot\epsilon_{+}(p_{5})] \nonumber \\
&  -\,\,\,(p_{3}-p_{4})\cdot(-p_{1}-p_{2}+p_{5})[(p_{2}-p_{1})\cdot\epsilon_{+}(p_{5})] \nonumber \\
&  -\,\,\,(p_{2}-p_{1})\cdot(p_{3}+p_{4}-p_{5})[(p_{3}-p_{4})\cdot\epsilon_{+}(p_{5})] ,\nonumber \\
n_{4}&= (p_{4}+p_{5})^{2}[(p_{2}-p_{3})\cdot\epsilon_{+}(p_{5})]+2(p_{2}-p_{3})\cdot(p_{1}-p_{4}-p_{5})
[p_{4}\cdot \epsilon_{+}(p_{5})], \nonumber \\
n_{5}&= -(p_{1}+p_{5})^{2}[(p_{3}-p_{4})\cdot \epsilon_{+}(p_{5})]-2(p_{3}-p_{4})\cdot(-p_{1}+p_{2}-p_{5})[p_{1}\cdot\epsilon_{+}(p_{5})], \nonumber \\
n_{6}&= -(p_{2}+p_{5})^{2}[(p_{4}-p_{1})\cdot\epsilon_{+}(p_{5})]-2(p_{4}-p_{1})\cdot(-p_{2}+p_{3}-p_{5})
[p_{2}\cdot\epsilon_{+}(p_{5})], \nonumber \\
n_{7}&= -(p_{2}+p_{5})^{2}[(p_{3}-p_{1})\cdot\epsilon_{+}(p_{5})]-2(p_{3}-p_{1})\cdot(-p_{2}+p_{4}-p_{5})
[p_{2}\cdot\epsilon_{+}(p_{5})], \nonumber \\
n_{8}&= (p_{2}-p_{4})\cdot(p_{1}+p_{3}-p_{5})[(p_{3}-p_{1})\cdot\epsilon_{+}(p_{5})] \nonumber \\
&  +\,\,\,(p_{3}-p_{1})\cdot(-p_{2}-p_{4}+p_{5})[(p_{2}-p_{4})\cdot\epsilon_{+}(p_{5})] \label{eq:numerators_indistinguisable}\\
&  +\,\,\,(p_{3}-p_{1})\cdot(p_{2}-p_{4})[(-p_{1}+p_{2}-p_{3}+p_{4})\cdot\epsilon_{+}(p_{5})], \nonumber\\
n_{9}&= -(p_{3}+p_{5})^{2}[(p_{4}-p_{2})\cdot\epsilon_{+}(p_{5})]-2(p_{4}-p_{2})\cdot(-p_{3}+p_{1}-p_{5})
[p_{3}\cdot\epsilon_{+}(p_{5})], \nonumber  \\
n_{10}&= -(p_{3}+p_{5})^{2}[(p_{4}-p_{1})\cdot\epsilon_{+}(p_{5})]-2(p_{4}-p_{1})\cdot(p_{2}-p_{3}-p_{5})
[p_{3}\cdot\epsilon_{+}(p_{5})], \nonumber \\
n_{11}&= -(p_{1}+p_{5})^{2}[(p_{2}-p_{4})\cdot\epsilon_{+}(p_{5})]-2(p_{2}-p_{4})\cdot(-p_{1}+p_{3}-p_{5})
[p_{1}\cdot\epsilon_{+}(p_{5})], \nonumber \\
n_{12} &= -
(p_{3}+p_{5})^{2}[(p_{2}-p_{1})\cdot\epsilon_{+}(p_{5})]
-2(p_{2}-p_{1})\cdot (-p_{3}+p_{4}-p_{5})[p_{3}\cdot\epsilon_{+}(p_{5})], \nonumber \\
n_{13}&= (p_{2}-p_{3})\cdot(p_{4}-p_{1})[(-p_{1}+p_{2}+p_{3}-p_{4})\cdot\epsilon(p_{5})] \nonumber \\
&  +\,\,\,(p_{4}-p_{1})\cdot(-p_{2}-p_{3}+p_{5})[(p_{2}-p_{3})\cdot\epsilon_{+}(p_{5})] \nonumber \\
&  +\,\,\,(p_{2}-p_{3})\cdot(p_{1}+p_{4}-p_{5})[(p_{4}-p_{1})\cdot\epsilon_{+}(p_{5})],\nonumber \\
n_{14}&= -(p_{2}+p_{5})^{2}[(p_{3}-p_{4})\cdot\epsilon_{+}(p_{5})]-2(p_{3}-p_{4})\cdot(-p_{2}+p_{1}-p_{5})
[p_{2}\cdot\epsilon_{+}(p_{5})], \nonumber \\
n_{15}&= (p_{4}+p_{5})^{2}[(p_{3}-p_{1})\cdot \epsilon_{+}(p_{5})]+2(p_{3}-p_{1})\cdot(p_{2}-p_{4}-p_{5})
[p_{4}\cdot\epsilon_{+}(p_{5})].  \nonumber 
\end{align}
We have assumed again that the emitted gluon has positive helicity. 

A long but straightforward evaluation of the amplitude in the planar limit gives the result
\begin{align}
\mathcal{A}_{5}\Bigg|_{\rm planar}&=-{2\sqrt{2}g^{3}\over \sqrt{s}}{(\zeta_{3}^{2}-\zeta_{3}\zeta_{4}+\zeta_{4}^{2}) \over (\zeta_{3}-\zeta_{4})
(1+\zeta_{3}\zeta_{4})}\left[-c_{2}{\zeta_{5}-\zeta_{3}\over \zeta_{3}}-c_{6}{\zeta_{4}-\zeta_{5}\over \zeta_{5}}+c_{7}{\zeta_{3}-\zeta_{5}\over \zeta_{5}}
\right.
\nonumber \\[0.2cm]
&-\left. c_{8}{\zeta_{3}-\zeta_{4}\over \zeta_{4}}+c_{11}{\zeta_{5}-\zeta_{4}\over \zeta_{4}}+c_{13}{\zeta_{4}-\zeta_{3}\over\zeta_{3}}
\right].
\end{align}
The prefactor $\zeta_{3}^{2}-\zeta_{3}\zeta_{4}+\zeta_{4}^{2}$ does not have real nontrivial zeros, corresponding to two complex straight lines in the $(\zeta_{3},\zeta_{4})$ plane. 
After multiplying
by $\zeta_{3}\zeta_{4}\zeta_{5}$, which does not introduce any spurious physical zeros, we arrive at the cubic
homogeneus equation
\begin{align}
c_{7}\zeta_{3}^{2}\zeta_{4}&-c_{8}\zeta_{3}^{2}\zeta_{5}-c_{6}\zeta_{3}\zeta_{4}^{2}+c_{11}\zeta_{3}\zeta_{5}^{2}
\nonumber \\[0.2cm]
&+(c_{2}+c_{6}-c_{7}+c_{8}-c_{11}-c_{13})\zeta_{3}\zeta_{4}\zeta_{5}+c_{13}\zeta_{4}^{2}\zeta_{5}-c_{2}\zeta_{4}\zeta_{5}^{2}=0.
\label{eq:zero_cond_gauge_case_indist}
\end{align}
Interestingly, the condition \eqref{eq:zero_cond_gauge_case_indist} 
for the existence of planar zeros in the scattering of 
two indistinguishable scalars with the emission of a gluon is identical to the one found for the five-gluon scattering amplitude in \cite{ours_pz}. The reader is referred to this reference for the analysis of the
curves for various gauge groups.

\section{String corrections to gauge theory planar zeros}
\label{sec:string}

It would be interesting to see how the planar zeros of (super) Yang-Mills theories get corrected when
considering ultraviolet completions such as open string theory. The full, $\alpha'$-exact disk amplitude
for the scattering of $n$ gauge bosons has a particularly simple structure \cite{MSS}
\begin{align}
\mathcal{A}_{n}(1\ldots n)_{\rm open}
=\sum_{\sigma\in S_{n-3}} F_{(1 \ldots n)}^{\sigma(2\ldots n-2)}(s_{ij};\alpha') A_{n}\Big(1,\sigma(2,\ldots,n-2),n-1,n\Big),
\label{eq:open_string_amp_general}
\end{align}
where $A_{n}(1,\ldots,n)$ is the field theory, color ordered gauge amplitude and 
$F_{(1\ldots n)}^{{\sigma(2\ldots n-2)}}(s_{ij};\alpha')$ are generalized Euler integrals over the Koba-Nielsen parameters
\begin{align}
F_{(1\ldots n)}^{\sigma(2\ldots n-2)}(s_{ij};\alpha')=(-1)^{n-3}\int\limits_{z_{i}<z_{i+1}}\left(\prod_{\ell=2}^{n-2} dz_{\ell}\right)\prod_{i<j}^{n-1}
|z_{i}-z_{j}|^{\alpha' s_{ij}} 
\left\{\prod_{k=2}^{n-2}\left(\sum_{m=1}^{k-1}{\alpha' s_{mk}\over z_{m}-z_{k}}\right)\right\}_{\sigma},
\label{eq:integralsF}
\end{align}
where the subindex $\sigma$ indicates that the permutation $\sigma\in S_{n-3}$ acts on all indices
inside the curly bracket. These integrals contain the whole $\alpha'$ dependence
of $\mathcal{A}_{n}(1,\ldots,n)_{\rm open}$ and can be seen as a dressing of the gauge theory amplitude 
to include the effect of the tower of massive string modes.

Similar to the case of Yang-Mills theories, a generic $n$-point open string amplitude can be expressed
in terms of a basis of $(n-3)!$ independent color ordered amplitudes \cite{BBDV}. It is convenient
to choose the basis
\begin{align}
\mathcal{A}_{n}\Big(1,\Pi_{a}(2,\ldots,n-2),n-1,n\Big)_{\rm open},
\end{align}
where $a=1,\ldots,(n-3)!$ and 
$\Pi_{a}$ denotes the elements of $S_{n-3}$. 
Then, Eq. \eqref{eq:open_string_amp_general} can be written in matrix form $\boldsymbol{\mathcal{A}}_{n}
=F\mathbf{A}_{n}$ as
\begin{align}
\begin{pmatrix}
\mathcal{A}_n(1,\Pi_1,n-1,n)\\
\vdots\\
\mathcal{A}_n(1,\Pi_{(n-3)!},n-1,n)
\end{pmatrix}
=
\begin{pmatrix}
F_{\Pi_1}^{\,\,\,\,\,\sigma_1} & \ldots & F_{\Pi_1}^{\,\,\,\,\,\sigma_{(n-3)!}} \\
\vdots &  & \vdots \\
F_{\Pi_{(n-3)!}}^{\,\,\,\,\,\sigma_1} & \ldots & F_{\Pi_{(n-3)!}}^{\,\,\,\,\,\sigma_{(n-3)!}}
\end{pmatrix}
\begin{pmatrix}
A_{n}(1,\sigma_1,n-1,n)\\
\vdots\\
A_{n}(1,\sigma_{(n-3)!},n-1,n)
\end{pmatrix},
\label{eq:matrix_form1}
\end{align}
where the shorthand notation $\Pi_{a}\equiv \Pi_{a}(2,\ldots,N-2)$ and 
$\sigma_{a}\equiv \sigma_{a}(2,\ldots,N-2)$ has been used. 

String corrections to field theory gauge amplitudes are obtained by 
expanding the integrals \eqref{eq:integralsF} in powers of $\alpha'$. The coefficients
of the series are expressed in terms of kinematic invariants and multiple zeta values (MZV).
Thus, the $(n-3)!\times (n-3)!$ matrix $F$ has the following expansion in powers of the inverse string
tension \cite{bss},
\begin{align}
F&=\mathbb{I}+\alpha'^{2}\zeta(2)P_{2}+\alpha'^{3}\zeta(3)M_{3}+\alpha'^{4}\zeta(2)^{2}P_{4}
+\alpha'^{5}\Big[\zeta(2)\zeta(3)P_{2}M_{3}+\zeta(5)M_{5}\Big]
\nonumber \\[0.2cm]
&+\alpha'^{6}\left[\zeta(2)^{3}P_{6}+{1\over 2}\zeta(3)^{2}M_{3}^{2}\right]
+\alpha'^{7}\Big[\zeta(7)M_{7}+\zeta(2)\zeta(5)P_{2}M_{7}+\zeta(2)^{2}\zeta(3)P_{4}M_{3}\Big]
+\ldots
\label{eq:F_expansion}
\end{align}
where 
\begin{align}
M_{2k+1}=F\Big|_{\zeta(2k+1)}, \hspace*{1cm} P_{2k}=F\Big|_{\zeta(2)^{k}}, 
\end{align}
with $P_{0}=\mathbb{I}$ and $M_{1}=0$. 
At order $\alpha'^{k}$, the matrix coefficient is a homogeneous function of degree $k$ in the
kinematic invariants $s_{ij}$. 

Let us particularize the analysis to the five-point function
\begin{align}
\begin{pmatrix}
\mathcal{A}_5(1,2,3,4,5)\\[0.2cm]
\mathcal{A}_5(1,3,2,4,5)
\end{pmatrix}
=
\begin{pmatrix}
F_{(12345)}^{\,\,\,\,(23)} & F_{(12345)}^{\,\,\,\,(32)} \\[0.2cm]
F_{(13245)}^{\,\,\,\,(23)} & F_{(13245)}^{\,\,\,\,(32)}
\end{pmatrix}
\begin{pmatrix}
A_5(1,2,3,4,5)\\[0.2cm]
A_5(1,3,2,4,5)
\end{pmatrix},
\label{eq:string_from_YM_amp5p}
\end{align}
where the matrix entries have the following expansion in powers of the string slope
\begin{align}
F_{(12345)}^{\,\,\,\,(23)}
&=1+\alpha'^{\,2}\zeta(2)(s_{12}s_{34}-s_{34}s_{45}-s_{12}s_{15})-\alpha'^{\,3}\zeta(3)\Big(s_{12}^2s_{34}+2s_{12}s_{23}s_{34}+s_{12}s_{34}^{2} \nonumber \\[0.2cm]
&-s_{34}^2s_{45}-s_{34}s_{45}^{2}-s_{12}^2s_{15}-s_{12}s_{15}^{2}\Big)+\mathcal{O}(\alpha'^{4}), 
\label{eq:expansionsF1}\\[0.2cm]
F_{(12345)}^{(32)}&=\alpha'^{\,2}\zeta(2)s_{13}s_{24}-\alpha'^{\,3}\zeta(3)s_{13}s_{24}\Big( s_{12}+s_{23}+s_{34}+s_{45}+s_{15}\Big)+\mathcal{O}(\alpha'^4),
\nonumber 
\end{align}
whereas 
\begin{align}
F_{(13245)}^{(23)}=F_{(12345)}^{(32)}\Big|_{2\leftrightarrow 3} 
\hspace*{1cm} \mbox{and} \hspace*{1cm} 
F_{(13245)}^{(32)}=F_{(12345)}^{(23)}\Big|_{2\leftrightarrow 3}.
\label{eq:expansionsF2}
\end{align}
Writing the kinematic invariants in \eqref{eq:expansionsF1} in terms of the stereographic coordinates, we see that the 
expansion parameter is $s\alpha'\ll 1$. This can be traced back to Eq. \eqref{eq:integralsF}, where all dependence on 
$\alpha'$ comes through the dimensionless combination $s_{ij}\alpha'=(s\alpha')f_{ij}(\zeta_{a})$, with $f_{ij}$ a function of the
stereographic coordinates.

We compute next the full color-dressed five-point disk amplitude. Following \cite{ours_pz},
we work in the Yang-Mills amplitude basis $A_{5}(1,\sigma(3,4,5),2)$, which means that we use 
the Jacobi identities to recast all color factors in terms of 
$\{c_{2},c_{6},c_{7},c_{8},c_{11},c_{13}\}$. Namely,
\begin{align}
\mathcal{A}_{5,\rm string}&=c_{2}\mathcal{A}_{5}(1,5,4,3,2)_{\rm open}+c_{6}\mathcal{A}_{5}(1,4,3,5,2)_{\rm open}+c_{7}\mathcal{A}_{5}(1,3,4,5,2)_{\rm open} \nonumber \\[0.2cm]
&+c_{8}\mathcal{A}_{5}(1,3,5,4,2)_{\rm open}+c_{11}\mathcal{A}_{5}(1,5,3,4,2)_{\rm open}+
c_{13}\mathcal{A}_{5}(1,4,5,3,2)_{\rm open}.
\end{align}
Using now Eq. \eqref{eq:string_from_YM_amp5p}, the full string amplitudes on the right-hand side of
this equation are expressed in terms of our basis of color-ordered Yang-Mills amplitudes as
\begin{align}
\mathcal{A}_{5,\rm string}&=\Big(c_{2}F^{(54)}_{(15432)}+c_{13}F^{(54)}_{(14532)}\Big)A_{5}(1,5,4,3,2)
+\Big(c_{6}F^{(43)}_{(14352)}+c_{7}F^{(43)}_{(13452)}\Big)A_{5}(1,4,3,5,2) \nonumber \\[0.2cm]
&+\Big(c_{6}F^{(34)}_{(14352)}+c_{7}F^{(34)}_{(13452)}\Big)A_{5}(1,3,4,5,2)+
\Big(c_{8}F^{(35)}_{(13542)}+c_{11}F^{(35)}_{(15342)}\Big)A_{5}(1,3,5,4,2) \nonumber \\[0.2cm]
&+\Big(c_{8}F^{(53)}_{(13542)}+c_{11}F^{(53)}_{(15342)}\Big)A_{5}(1,5,4,3,2) 
+\Big(c_{2}F^{(45)}_{(15432)}+c_{13}F^{(45)}_{(14532)}\Big)A_{5}(1,4,5,3,2).
\end{align}
Finally, we use the expressions for the color subamplitudes given by the Parke-Taylor 
formula\footnote{As in \cite{ours_pz}, we consider MHV amplitudes with helicities
$(1^{-},2^{-},3^{+},4^{+},5^{+})$.} \cite{parke_taylor}
and implement
the expansions \eqref{eq:expansionsF1} and \eqref{eq:expansionsF2}. Using the stereographic coordinates
defined in \eqref{eq:momentum_parametrization_in} and \eqref{eq:momentum_parametrization_out}, we arrive at the
final expression for the five-point disk amplitude at order $\alpha'^{\,3}$ in the planar limit:
\begin{align}
\mathcal{A}_{5,\rm string}\Big|_{\rm planar}&={i(\zeta_{3}-\zeta_{4})(\zeta_{3}-\zeta_{5})(\zeta_{4}-\zeta_{5})\over \sqrt{s}
\zeta_{3}\zeta_{4}\zeta_{5}(1+\zeta_{3}\zeta_{4})(1+\zeta_{3}\zeta_{5})(1+\zeta_{4}\zeta_{5})}\left[
A_{5}^{(0)}+{(s\alpha')^{2}\zeta(2)A_{5}^{(2)}\over (\zeta_{3}-\zeta_{4})(\zeta_{3}-\zeta_{5})
(\zeta_{4}-\zeta_{5})}
\right. \nonumber \\[0.2cm]
&\left.+{(s\alpha')^{3}\zeta(3)A_{5}^{(3)}\over (\zeta_{3}-\zeta_{4})^{2}(\zeta_{3}-\zeta_{5})^{2}
(\zeta_{4}-\zeta_{5})^{2}}+\mathcal{O}\Big((s\alpha')^{4}\Big)\right].
\label{eq:disk_planar_amp}
\end{align}

The coefficient $A_{5}^{(0)}$ is the cubic homogeneous polynomial determining the 
planar zeros of the five-gluon amplitude \cite{ours_pz} 
\begin{align}
A_{5}^{(0)}(\zeta_{3},\zeta_{4},\zeta_{5})
&=c_{7}\zeta_{3}^{2}\zeta_{4}-c_{8}\zeta_{3}^{2}\zeta_{5}-c_{6}\zeta_{3}\zeta_{4}^{2}+c_{11}\zeta_{3}\zeta_{5}^{2}
\nonumber \\[0.2cm]
&+(c_{2}+c_{6}-c_{7}+c_{8}-c_{11}+c_{13})\zeta_{3}\zeta_{4}\zeta_{5}+c_{13}\zeta_{4}^{2}\zeta_{5}
-c_{2}\zeta_{4}\zeta_{5}^{2}.
\end{align}
However, the $\alpha'^{\,2}$ and $\alpha'^{\,3}$ coefficients $A_{5}^{(2)}$ and $A_{5}^{(3)}$ are respectively degree 10
and 15, {\em nonhomogeneus} polynomials whose explicit expressions
are given in Eqs. \eqref{eq:A_5^(2)} and \eqref{eq:A_5^(3)} of the Appendix.
Thus, $\alpha'$ corrections destroy the projective properties of the loci of planar zeros found in \cite{ours_pz}. Interestingly, when considering the scattering
of two gluons in a singlet state
\begin{align}
c_{2}=c_{6}=-c_{7}=c_{8}=-c_{11}=-c_{13}=-f^{a_{3}a_{4}a_{5}},
\end{align}
the equation $A_{5}^{(2)}=0$ becomes a homogeneous polynomial
\begin{align}
\zeta_{3}\zeta_{4}\zeta_{5}(\zeta_{3}-\zeta_{4})(\zeta_{3}-\zeta_{5})(\zeta_{4}-\zeta_{5})=0.
\end{align}
However, the zeros of this equation all lie at unphysical values of the stereographic coordinates for which
either the amplitude or the energy of at least one of the outgoing particles diverges. 

\section{Gravitational amplitudes}
\label{sec:grav_scalars}

One of the results of Ref. \cite{ours_pz} is that the planar, MHV five-point graviton amplitude is identically zero. This fact can be seen
as a consequence of the theorem proved in \cite{huang_johansson}, stating the vanishing of all helicity violating amplitudes
in three dimensions. Indeed, at the level of the tree amplitude, the graviton couplings are of the form 
$p_{i}\cdot\varepsilon_{k}\cdot p_{j}$ with $i,j\neq k$, so
imposing planarity decouples the graviton polarization normal to the plane. This renders the scattering effectively three-dimensional and, 
as a consequence, the planar MHV amplitude is equal to zero. 

In this section we are going to explore other gravitational amplitudes involving scalar particles minimally coupled to gravity. We begin with the
scattering of two distinguishable scalars with graviton emission
\begin{align}
\Phi(p_{1})+\Phi'(p_{2})\longrightarrow \Phi(p_{3})+\Phi'(p_{4})+G(p_{5},\varepsilon).
\label{eq:grav_dist_scalars_proc}
\end{align}
The tree-level amplitude was computed in Ref. \cite{sv_sc_vm1} using the Feynman rules for a scalar theory coupled to gravity. 
Using the Sudakov decomposition, 
\begin{align}
k_{1}&\equiv -p_{1}-p_{3}=\alpha_{1}p_{1}+\beta_{1}p_{2}+k_{1,\perp}, \nonumber \\[0.2cm]
k_{2}&\equiv -p_{2}-p_{4}=\alpha_{2}p_{1}+\beta_{2}p_{2}+k_{2,\perp},
\end{align}
the amplitude has the tensor structure
\begin{align}
\mathcal{M}&=\left({\kappa\over 2}\right)^{3}\Bigg\{(K_{\perp}\cdot\varepsilon\cdot K_{\perp})A_{KK}+\Big[(K_{\perp}\cdot\varepsilon\cdot p_{1})
+(p_{1}\cdot \varepsilon\cdot K_{\perp})\Big]A_{K1} \nonumber \\[0.2cm]
&+\Big[(K_{\perp}\cdot\varepsilon\cdot p_{2})
+(p_{2}\cdot \varepsilon\cdot K_{\perp})\Big]A_{K2}+(p_{1}\cdot\varepsilon\cdot p_{1})A_{11}+(p_{2}\cdot\varepsilon\cdot p_{2})A_{22}
\label{eq:grav_amp_dis_sc_full1}\\[0.2cm]
&+\Big[(p_{1}\cdot\varepsilon\cdot p_{2})
+(p_{2}\cdot \varepsilon\cdot p_{1})\Big]A_{12}\Bigg\},
\nonumber
\end{align}
where $K_{\perp}\equiv k_{1,\perp}+k_{2,\perp}$ and the coefficients $A_{i}$ are rational functions of the Sudakov parameters $\alpha_{i}$, 
$\beta_{i}$. The tensor structure of the amplitude shows again how, once the planar limit $\zeta_{i}=\overline{\zeta}_{i}$ is taken, 
the polarizations outside the interaction plane decouple and the amplitude becomes effectively three-dimensional.  
In this limit, the Sudakov parameters take the following form in terms of the stereographic coordinates:
\begin{align}
\alpha_{1}&\equiv {p_{2}\cdot (p_{1}+p_{3})\over p_{1}\cdot p_{2}}={\zeta_{4}\zeta_{5}(1-\zeta_{3}^{2})-\zeta_{3}(\zeta_{4}+\zeta_{5})
\over (\zeta_{3}-\zeta_{4})(\zeta_{3}-\zeta_{5})}, \nonumber \\[0.2cm]
\beta_{1}&\equiv {p_{1}\cdot (p_{1}+p_{3}) \over p_{1}\cdot p_{2}}= -{1+\zeta_{4}\zeta_{5}\over (\zeta_{3}-\zeta_{4})(\zeta_{3}-\zeta_{5})}, \nonumber \\[0.2cm]
\alpha_{2}&\equiv -{p_{2}\cdot (p_{2}+p_{4})\over p_{1}\cdot p_{2}}=-{\zeta_{4}^{2}(1+\zeta_{3}\zeta_{5})\over (\zeta_{3}-\zeta_{4})(\zeta_{4}-\zeta_{5})}, \\[0.2cm]
\beta_{2}&\equiv -{p_{1}\cdot (p_{2}+p_{4})\over p_{1}\cdot p_{2}}={-1+\zeta_{4}(-\zeta_{3}+\zeta_{4}-\zeta_{5})\over (\zeta_{3}-\zeta_{4})(\zeta_{4}-\zeta_{5})},
\nonumber
\end{align}
while the graviton polarization tensor is taken to be $\varepsilon_{\pm}=\epsilon_{\pm}\otimes\epsilon_{\pm}$, with $\epsilon_{\pm}$ defined by 
\eqref{eq:gluon_pol}.
Using the explicit expression for the coefficients in \eqref{eq:grav_amp_dis_sc_full1} given in \cite{sv_sc_vm1}, we find that the planar amplitude
vanishes identically
\begin{align}
\mathcal{M}\Big|_{\rm planar}=0.
\end{align}

It was found in \cite{sv_sc_vm1} that this gravitational amplitude can be split into two gauge invariant subamplitudes, 
$\mathcal{M}=\mathcal{M}_{\uparrow}+\mathcal{M}_{\downarrow}$, where each term can be written in terms of an effective, nonlocal
vertex. In the planar limit, these subamplitudes are individually nonzero and take a specially simple form
\begin{align}
\mathcal{M}_{\uparrow}\Big|_{\rm planar}
=-\mathcal{M}_{\downarrow}\Big|_{\rm planar}
=\left({\kappa s\over 4}\right){\zeta_{3}(1+\zeta_{3}\zeta_{5})(1+\zeta_{4}\zeta_{5})\over
(1+\zeta_{3}\zeta_{4})(\zeta_{4}-\zeta_{5}+\zeta_{3}\zeta_{4}\zeta_{5}-\zeta_{4}\zeta_{5}^{2})}.
\end{align} 

The gravitational amplitude \eqref{eq:grav_amp_dis_sc_full1} 
cannot be retrieved using the double-copy BCJ construction \cite{BCJ} from the
gauge scattering amplitude of two distinct scalars with a 
gluon emission \cite{sv_sc_vm2}. Despite this, the putative gravitational amplitude obtained from the double-copy
of the gauge amplitude \eqref{eq:sQCD_amp}, with denominators satisfying color-kinematics duality, identically 
vanishes in the planar limit.  

A similar result is obtained for the gravitational scattering of two indistinguishable scalars. In this case, the amplitude
can be obtained by double copy from the gauge scattering of adjoint identical scalars given in Eq. \eqref{eq:amplitude_ind_scalars_general} 
using color-kinematics duality \cite{jsvscvm},
\begin{align}
\mathcal{M}_{5}&=\left({\kappa\over 2}\right)^{3}\left({n_{1}^{2}\over s_{12}s_{45}}+{n_{2}^{2}\over s_{23}s_{15}}+{n_{3}^{2}\over s_{34}s_{12}}
+{n_{4}^{2}\over s_{45}s_{23}}+{n_{5}^{2}\over s_{15}s_{34}}+{n_{6}^{2}\over s_{14}s_{25}}
+{n_{7}^{2}\over s_{13}s_{25}}+{n_{8}^{2}\over s_{24}s_{13}} \right. \nonumber \\[0.2cm]
&+ \left.{n_{9}^{2}\over s_{35}s_{24}}+{n_{10}^{2}\over s_{14}s_{35}}+{n_{11}^{2}\over s_{15}s_{24}}
+{n_{12}^{2}\over s_{12}s_{35}}+{n_{13}^{2}\over s_{23}s_{14}}+{n_{14}^{2}\over s_{25}s_{34}}
+{n_{15}^{2}\over s_{13}s_{45}}\right),
\end{align}
where the numerators are the ones given in Eq. \eqref{eq:numerators_indistinguisable}. In fact, the cancellation of this amplitude in the planar limit
can be seen to happen by a mechanism similar to the one found in \cite{ours_pz} for the pure gravitational case. Indeed, 
replacing the color factors by the corresponding numerators in the condition for the gauge planar zeros \eqref{eq:zero_cond_gauge_case_indist},
we find the following condition for the existence of planar zeros
\begin{align}
n_{7}\zeta_{3}^{2}\zeta_{4}-n_{8}\zeta_{3}^{2}\zeta_{5}-n_{6}\zeta_{3}\zeta_{4}^{2}+n_{11}\zeta_{3}\zeta_{5}^{2}
\hspace*{6cm}\nonumber \\[0.2cm]
+(n_{2}+n_{6}-n_{7}+n_{8}-n_{11}-n_{13})\zeta_{3}\zeta_{4}\zeta_{5}+n_{13}\zeta_{4}^{2}\zeta_{5}-n_{2}\zeta_{4}\zeta_{5}^{2}=0.
\label{eq:amplitude_ind_scalars_gravity_BCJ}
\end{align}

In the planar limit (i.e., real stereographic coordinates), the relevant numerators have the following form 
\begin{align}
n_{2}&= {s^{3\over 2}\over \sqrt{2}(\zeta_{3}-\zeta_{4})(\zeta_{3}-\zeta_{5})(\zeta_{4}-\zeta_{5})(1+\zeta_{5}^{2})}
\Big(1+\zeta_{3}\zeta_{4}-2\zeta_{3}\zeta_{5}+5\zeta_{4}\zeta_{5}-2\zeta_{3}^{2}\zeta_{5}^{2} \nonumber \\[0.2cm]
&+\zeta_{3}\zeta_{4}\zeta_{5}^{2}
+4\zeta_{4}^{2}\zeta_{5}^{2}-\zeta_{3}^{2}\zeta_{4}^{2}\zeta_{5}^{2}-2\zeta_{3}^{2}\zeta_{4}\zeta_{5}^{3}+4\zeta_{3}\zeta_{4}^{2}\zeta_{5}^{3}\Big), \nonumber
\\[0.2cm]
n_{6}&={s^{3\over 2}\zeta_{5}\over \sqrt{2}(\zeta_{3}-\zeta_{4})(\zeta_{3}-\zeta_{5})(\zeta_{4}-\zeta_{5})(1+\zeta_{5}^{2})}
\Big(-2\zeta_{3}+4\zeta_{4}-\zeta_{5}-2\zeta_{3}^{2}\zeta_{5}+\zeta_{3}\zeta_{4}\zeta_{5} \nonumber \\[0.2cm]
&+4\zeta_{4}^{2}\zeta_{5}-2\zeta_{3}^{2}\zeta_{4}\zeta_{5}^{2}+4\zeta_{3}\zeta_{4}^{2}\zeta_{5}^{2}
+\zeta_{3}\zeta_{4}\zeta_{5}^{3}+\zeta_{3}^{2}\zeta_{4}^{2}\zeta_{5}^{3}\Big), \nonumber \\[0.2cm]
n_{7}&=-{s^{3\over 2}\zeta_{5}\over \sqrt{2}(\zeta_{3}-\zeta_{4})(\zeta_{3}-\zeta_{5})(\zeta_{4}-\zeta_{5})(1+\zeta_{5}^{2})}
(4\zeta_{3}-2\zeta_{4}-\zeta_{5}+4\zeta_{3}^{2}\zeta_{5}+\zeta_{3}\zeta_{4}\zeta_{5} \nonumber \\[0.2cm]
&-2\zeta_{4}^{2}\zeta_{5}+4\zeta_{3}^{2}\zeta_{4}\zeta_{5}^{2}-2\zeta_{3}\zeta_{4}^{2}\zeta_{5}^{2}+\zeta_{3}\zeta_{4}\zeta_{5}^{3}
+\zeta_{3}^{2}\zeta_{4}^{2}\zeta_{5}^{3}\Big), \nonumber \\[0.2cm]
n_{8}&={s^{3\over 2}\over \sqrt{2}(\zeta_{3}-\zeta_{4})^{2}(\zeta_{3}-\zeta_{5})(\zeta_{4}-\zeta_{5})(1+\zeta_{5}^{2})}
\Big(\zeta_{4}-\zeta_{5}-\zeta_{3}^{3}\zeta_{4}^{3}\zeta_{5}^3-4\zeta_{3}^{3}\zeta_{4}^{2}\zeta_{5}^{2}  \\[0.2cm]
&-4\zeta_{3}^{3}\zeta_{4}\zeta_{5}
+\zeta_{3}^{2}\zeta_{4}^{3}\zeta_{5}^{4}+\zeta_{3}^{2}\zeta_{4}^{3}\zeta_{5}^{2}+8\zeta_{3}^{2}\zeta_{4}^{2}\zeta_{5}^{3}
-\zeta_{3}^{2}\zeta_{4}^{2}\zeta_{5}-4\zeta_{3}^{2}\zeta_{4}\zeta_{5}^4+8\zeta_{3}^{2}\zeta_{4}\zeta_{5}^2 \nonumber \\[0.2cm]
&-4\zeta_{3}^{2}\zeta_{4}-4\zeta_{3}^{2}\zeta_{5}^{3}-4\zeta_{3}\zeta_{4}^{3}\zeta_{5}^{3}
+2\zeta_{3}\zeta_{4}^{3}\zeta_{5}+3\zeta_{3}\zeta_{4}^{2}\zeta_{5}^{4}+2\zeta_{3}\zeta_{4}^{2}\zeta_{5}^{2}
+3\zeta_{3}\zeta_{4}^{2} \nonumber \\[0.2cm]
&-\zeta_{3}\zeta_{4}\zeta_{5}^{3}+8\zeta_{3}\zeta_{4}\zeta_{5}
-4\zeta_{3}\zeta_{5}^{2}-4\zeta_{4}^{3}\zeta_{5}^{2}+2\zeta_{4}^{2}\zeta_{5}^{3}-4\zeta_{4}^{2}\zeta_{5}
+\zeta_{4}\zeta_{5}^{2}\Big),
\nonumber \\[0.2cm]
n_{11}&={s^{3\over 2}\over \sqrt{2}(\zeta_{3}-\zeta_{4})(\zeta_{3}-\zeta_{5})(\zeta_{4}-\zeta_{5})(1+\zeta_{5}^{2})}
\Big(-1-\zeta_{3}\zeta_{4}-4\zeta_{3}\zeta_{5}-4\zeta_{3}^{2}\zeta_{5}^{2} \nonumber \\[0.2cm]
&-\zeta_{3}\zeta_{4}\zeta_{5}^{2}+2\zeta_{4}^{2}\zeta_{5}^{2}+\zeta_{3}^{2}\zeta_{4}^{2}\zeta_{5}^{2}-4\zeta_{3}^{2}\zeta_{4}\zeta_{5}^{3}
+2\zeta_{3}\zeta_{4}^{2}\zeta_{5}^{3}\Big), \nonumber \\[0.2cm]
n_{13}&={s^{3\over 2}\over \sqrt{2}(\zeta_{3}-\zeta_{4})^{2}(\zeta_{3}-\zeta_{5})(\zeta_{4}-\zeta_{5})(1+\zeta_{5}^{2})}
\Big(\zeta_{3}-\zeta_{5}-\zeta_{3}^{3}\zeta_{4}^{3}\zeta_{5}^{3}+\zeta_{3}^{3}\zeta_{4}^{2}\zeta_{5}^{4}\nonumber \\[0.2cm]
&+\zeta_{3}^{3}\zeta_{4}^{2}\zeta_{5}^{2}
-4\zeta_{3}^{3}\zeta_{4}\zeta_{5}^{3}+2\zeta_{3}^{3}\zeta_{4}\zeta_{5}-4\zeta_{3}^{3}\zeta_{5}^{2}-4\zeta_{3}^{2}\zeta_{4}^{3}\zeta_{5}^{2}
+8\zeta_{3}^{2}\zeta_{4}^{2}\zeta_{5}^{3}-\zeta_{3}^{2}\zeta_{4}^{2}\zeta_{5} \nonumber \\[0.2cm]
&+3\zeta_{3}^{2}\zeta_{4}\zeta_{5}^{4}
+2\zeta_{3}^{2}\zeta_{4}\zeta_{5}^{2}+3\zeta_{3}^{2}\zeta_{4}+2\zeta_{3}^{2}\zeta_{5}^{3}
-4\zeta_{3}^{2}\zeta_{5}-4\zeta_{3}\zeta_{4}^{3}\zeta_{5}-4\zeta_{3}\zeta_{4}^{2}\zeta_{5}^{4} \nonumber \\[0.2cm]
&+8\zeta_{3}\zeta_{4}^{2}\zeta_{5}^{2}-4\zeta_{3}\zeta_{4}^{2}-\zeta_{3}\zeta_{4}\zeta_{5}^{3}
+8\zeta_{3}\zeta_{4}\zeta_{5}+\zeta_{3}\zeta_{5}^{2}-4\zeta_{4}^{2}\zeta_{5}^{3}-4\zeta_{4}\zeta_{5}^{2}\Big).
\nonumber
\end{align}
Substituting these values in Eq. \eqref{eq:amplitude_ind_scalars_gravity_BCJ}, we conclude that the condition for the existence of planar zeros
is identically satisfied for any kinematic configuration. 
Since the numerators now are far more complicated than the ones for gluon scattering \cite{ours_pz}, 
the cancellation taking place is less trivial.

Since the scalar gravitational amplitudes studied above do not preserve helicity, the fact that they are zero 
in the planar limit 
is also a consequence of the vanishing of all 
helicity-violating supergravity amplitudes when reduced to
three dimensions \cite{huang_johansson} (see also \cite{elvang_huang}). 
Indeed, the gauge amplitude for indistinguishable adjoint scalars \eqref{eq:amplitude_ind_scalars_general} 
can be embedded in a $\mathcal{N}=2$ super Yang-Mills theory \cite{jsvscvm}. Thus, the corresponding 
double copy can be thought of as a scattering amplitude  
in $\mathcal{N}=4$ supergravity \cite{cgjr}. In the case of the gravitational scattering of two distinct scalars 
\eqref{eq:grav_dist_scalars_proc}, on the other hand, the theory can be also embedded in a four-dimensional 
supergravity theory, such as the ones studied in \cite{cgjr}. Both amplitudes vanish in the planar limit, where
the dynamics becomes effectively three-dimensional.
 
In the case of graviton MHV amplitudes, their vanishing in the planar limit follows from the explicit expression of the $n$-graviton amplitude
\cite{MHVgrav_amp}
\begin{align}
M^{\rm MHV}_{n}&=\sum_{P(1,\ldots,n-3)}{1\over \langle n\, n-2\rangle\langle n-2\, n-1\rangle \langle n-1\, n\rangle}
{1\over \langle 12\rangle\ldots \langle n 1\rangle} \nonumber \\[0.2cm]
&\times \prod_{k=1}^{n-3}{[k|p_{k+1}+\ldots+p_{n-2}|n-1\rangle\over
\langle k\, n-1\rangle},
\label{eq:MHVgravamp}
\end{align}
where the sum runs over all permutation of the labels $1,\ldots,n-3$ and the notation
\begin{align}
[a|p_{k_{1}}+\ldots+p_{k_{n}}|b\rangle\equiv [ak_{1}]\langle k_{1}b\rangle+\ldots+
[ak_{n}]\langle k_{n}b\rangle,
\end{align}
has been used. 
Using this expression, we have explicitly checked
that
\begin{align}
M_{n}^{\rm MHV}\Big|_{\rm planar}=0, \hspace*{1cm} \mbox{for $n=5,6,7$, and $8$},
\end{align}
as expected. 

For the scattering of four gravitons, Eq. \eqref{eq:MHVgravamp} gives the helicity preserving amplitude
$M_{4}(1^{+},2^{+},3^{-},4^{-})$. Notice that the four-point amplitude is always planar, and the result obtained
by applying \eqref{eq:MHVgravamp} is however different from zero
\begin{align}
M_{4}\Big|_{\rm planar}\equiv M_{4}=s\left({1+\zeta_{3}^{2}\over \zeta_{3}}\right)^{2},
\end{align}
where we have chosen coordinates such that the process takes place on the plane $y=0$ (i.e., $\zeta_{3}
=-\zeta_{4}^{-1}\in 
\mathbb{R}$). Another scattering amplitude whose vanishing in the planar limit is 
not implied by the results of
\cite{huang_johansson} is the six-graviton, helicity preserving amplitude $M_{6}(1^{+},2^{-},3^{-},
4^{-},5^{+},6^{+})$. This can be computed starting with the six-gluon, helicity preserving amplitude
\cite{henn_plefka},
\begin{align}
A_{6}(1^{+},2^{-},3^{-},4^{-},5^{+},6^{+})&={\langle 4|p_{2}+p_{3}|1]^{3}
\over (p_{1}+p_{2}+p_{3})^{2}[12][23]\langle 45\rangle \langle 56\rangle \langle 6|p_{1}+p_{2}|3]}
\nonumber \\[0.2cm]
&+{\langle 2|p_{1}+p_{6}|5]^{3}\over (p_{1}+p_{2}+p_{6})^{2}[34][45]\langle 61\rangle\langle 12\rangle
\langle 6|p_{1}+p_{2}|3]},
\label{eq:six_grav_amp}
\end{align}
and applying the KLT formula
\begin{align}
M_{6}=-\kappa^{4}\boldsymbol{A}_{6}^{T}S_{0}\boldsymbol{A}_{6}
\end{align}
where $S_{0}$ is the field theory KLT kernel introduced in Eq. \eqref{eq:FSF=S0svF}. 

\begin{figure}[t]
\centerline{\includegraphics[scale=0.9]{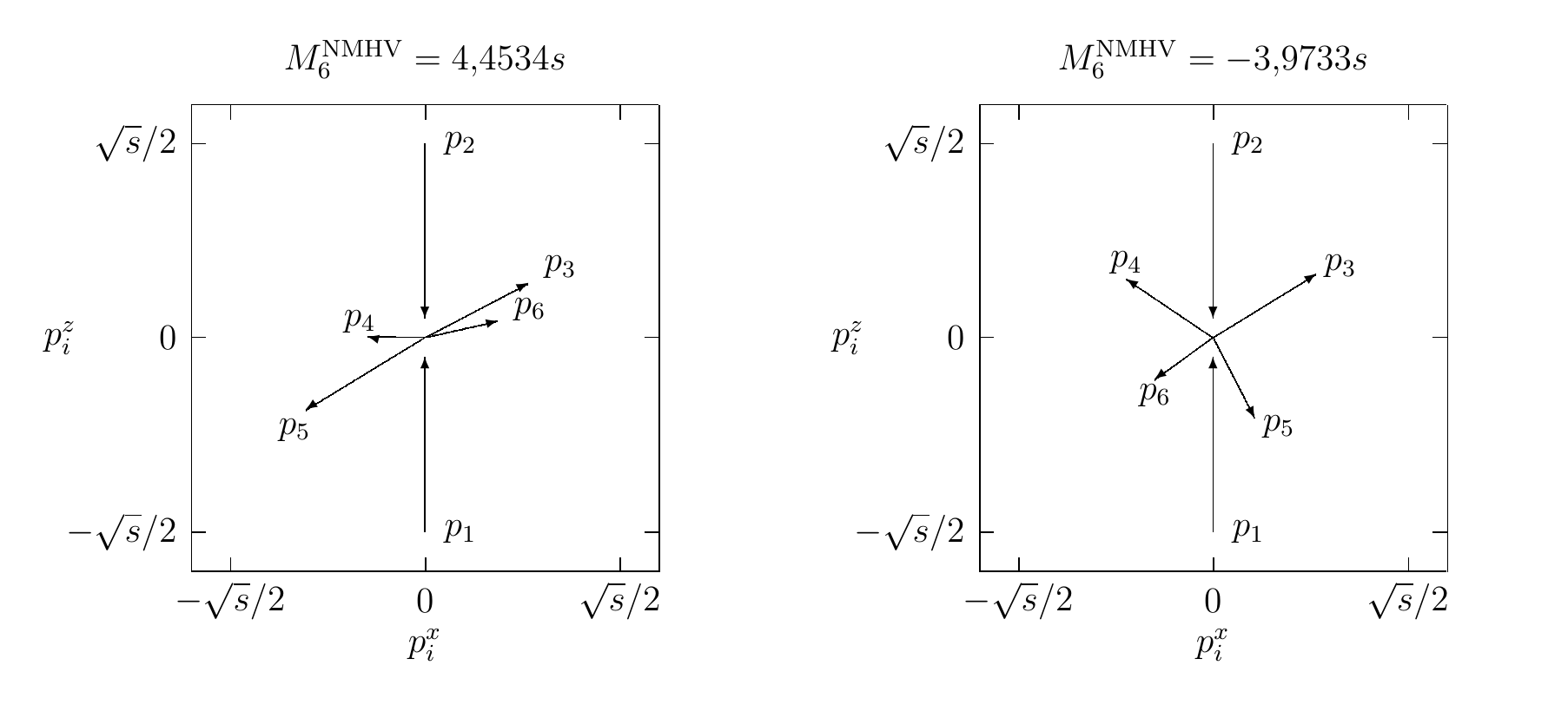}}
\caption[]{Two kinematic configurations for which the planar, helicity preserving six-graviton amplitude
in Eq. \eqref{eq:six_grav_amp} is nonzero.}
\label{fig:nonzero_grav_six}
\end{figure}

The explicit expression for the amplitude $M_{6}(1^{+},2^{-},3^{-},4^{-},5^{+},6^{+})$ in the planar limit 
in terms of the stereographic coordinates is very cumbersome and will not be given here. However, it can be seen that this amplitude does not vanish. In Fig. \ref{fig:nonzero_grav_six} we have depicted two kinematic {\em planar} configurations for which a calculation of the
tree-level amplitude gives a nonzero result.

\section{String corrections to graviton planar scattering}
\label{sec:closed_strings}

String graviton amplitudes on the sphere can be written in terms of disk amplitudes of gauge bosons using the Kawai-Lewellen-Tye 
(KLT) relations \cite{KLT}. A general expression for the $n$-gravity amplitude reads \cite{BBDSV}
\begin{align}
\mathcal{M}_{n}&=(-1)^{n-3}\kappa^{n-2}\sum_{\sigma,\rho\in S_{n-3}}\mathcal{A}_{n}\Big(1,\sigma(2,\ldots,n-2),n-1,n\Big) \nonumber \\[0.2cm]
&\times \mathcal{S}[\rho|\sigma]_{1}\widetilde{\mathcal{A}}_{n}\Big(1,\rho(2,\ldots,n-2),n,n-1\Big),
\label{eq:KLT_generalized}
\end{align}
where the two gauge copies differ by the ordering of the last two entries. The momentum kernel $\mathcal{S}[\rho|\sigma]_{1}$ has the form
\begin{align}
\mathcal{S}[\rho|\sigma]_{1}&\equiv \mathcal{S}[\rho(2,\ldots,n-2)|\sigma(2,\ldots,n-2)]_{1}
\nonumber \\[0.2cm]
&=\left({2\over \pi\alpha'}\right)^{n-3}\prod_{j=2}^{n-2}\sin\left[{\pi\alpha'\over 2}\left(s_{1j_{p}}+\sum_{k=2}^{j-1}
\theta(j_{\rho},k_{\rho})s_{j_{\rho}k_{\rho}}\right)\right],
\label{eq:KLT_kernel_full}
\end{align}
where the symbol $\theta(j_{\rho},k_{\rho})$ equals $1$ if the legs $j_{\rho}$ and $k_{\rho}$ keep the same order in the sets
$\rho(2,\ldots,n-2)$ and $\sigma(2,\ldots,n-2)$, and $0$ otherwise.   

The generalized KLT relations \eqref{eq:KLT_generalized} can be recast in matrix form as \cite{svp2}
\begin{align}
\mathcal{M}_{n}=(-1)^{n-3}\kappa^{n-2}\widetilde{\boldsymbol{\mathcal{A}}}_{n}^{T}{\mathcal{S}}\boldsymbol{\mathcal{A}}_{n}
=(-1)^{n-3}\kappa^{n-2}\boldsymbol{\mathcal{A}}_{n}^{T}{\mathcal{S}}_{0}\boldsymbol{\mathcal{A}}_{n},
\end{align}
where in the second identity we have changed the basis of the first-copy amplitudes to express them in terms of 
of the basis $\boldsymbol{\mathcal{A}}_{n}$ used in Eq. \eqref{eq:matrix_form1}. Using now this same equation, we
can express the string graviton amplitude in terms of field theory gauge amplitudes as
\begin{align}
\mathcal{M}_{n}=(-1)^{n-3}\kappa^{n-2}\boldsymbol{A}_{n}^{T}F^{T}\mathcal{S}_{0}F\boldsymbol{A}_{n}.
\end{align}
The single-valued projection \cite{brown,stieberger_taylor,svp,svp2} allows a further simplification of this relation. It projects the MZVs appearing in 
the expansion of the matrix $F$ in \eqref{eq:F_expansion} to a subclass ${\rm sv}(F)$, 
called the single-valued MZVs, which exactly reproduces the
closed string $\alpha'$ expansion
\begin{align}
F^{T}\mathcal{S}_{0}F=S_{0}\,{\rm sv}(F),
\label{eq:FSF=S0svF}
\end{align}
where $S_{0}$ is the field theory limit ($\alpha'\rightarrow 0$) KLT kernel in the basis $\boldsymbol{\mathcal{A}}_{n}$.
The action of the single-valued projection on the MZVs is given by
\begin{align}
{\rm sv}[\zeta(2)]&=0, \nonumber \\[0.2cm]
{\rm sv}[\zeta(2n+1)]&=2\zeta(2n+1), \hspace*{1cm} \mbox{for $n \geq 1$}. 
\label{eq:svzeta}
\end{align}
With this, the string amplitude takes the form
\begin{align}
\mathcal{M}_{n}=(-1)^{n-3}\kappa^{n-2}\boldsymbol{A}_{n}^{T}S_{0}\,{\rm sv}(F)\boldsymbol{A}_{n}.
\label{eq:graviton_amp_closed_st}
\end{align}
Incidentally, dropping the term ${\rm sv}(F)$ in the previous expression we retrieve the KLT expression of the field theory
graviton amplitude.

We particularize our analysis to the five-point amplitude
\begin{align}
\mathcal{M}_{5}=\kappa^{3}\boldsymbol{A}_{5}^{T}S_{0}\,{\rm sv}(F)\boldsymbol{A}_{5},
\label{eq:M5_string_full}
\end{align}
where, from \eqref{eq:F_expansion}, we have
\begin{align}
{\rm sv}(F)&=\mathbb{I}+2\left({\alpha'\over 4}\right)^{3}\zeta(3)M_{3}
+2\left({\alpha'\over 4}\right)^{5}\zeta(5)M_{5}
\nonumber \\[0.2cm]
&+2\left({\alpha'\over 4}\right)^{6}\zeta(3)^{2}M_{3}^{2}
+2\left({\alpha'\over 4}\right)^{7}\zeta(7)M_{7}
+\ldots
\label{eq:sv(F)exp}
\end{align}
In order to get the closed string expression, 
we have to perform the rescaling $\alpha'\rightarrow \alpha'/4$, as explained in \cite{stieberger_taylor}.
Notice that the single-valued projection \eqref{eq:svzeta} eliminates many terms in the $\alpha'$-expansion of $F$. 
Plugging \eqref{eq:sv(F)exp} into Eq. \eqref{eq:graviton_amp_closed_st}, we see how
the first term gives, via the KLT relations, the field theory gravity amplitude, while the second one
corresponds to the first nonvanishing string correction. 
The entries of the matrix $M_{3}$ can be read from Eqs. \eqref{eq:expansionsF1} and \eqref{eq:expansionsF2}.
The matrix $S_{0}$ is given by $S_{0}=K^{T}S$, where $S$ is the $\alpha'\rightarrow 0$ limit of the KLT kernel in Eq. \eqref{eq:KLT_kernel_full} 
and $K$ implements the change of basis $\widetilde{\boldsymbol{A}}_{5}=K\boldsymbol{A}_{5}$ in the first copy. 
Using the Kleiss-Kuijf and BCJ relations, this matrix is given by
\begin{align}
K=\left(
\begin{array}{cc}
{s_{34}(s_{35}-s_{24})\over s_{14}s_{35}} & {s_{13}s_{24}\over s_{14}s_{35}} \\[0.2cm]
{s_{12}s_{34}\over s_{14}s_{25}} & {s_{24}(s_{25}-s_{34})\over s_{14}s_{25}}
\end{array}
\right).
\end{align}

Imposing the planarity condition in the stereographic coordinates, $\zeta_{a}\in \mathbb{R}$, we confirm the result of 
\cite{ours_pz}
\begin{align}
\kappa^{3}\boldsymbol{A}_{5}^{T}S_{0}\boldsymbol{A}_{5}\Bigg|_{\rm planar}=0. 
\end{align} 
However, a first nonvanishing string correction survives the planar limit,
\begin{align}
\mathcal{M}_{5}\Bigg|_{\rm planar}={3\zeta(3)\over 32}\alpha'^{3}\kappa^{3}s^{4}+\mathcal{O}(\alpha'^{5}). 
\end{align}
This term is independent of the directions of the final states and is never zero. Using the expansion
\eqref{eq:F_expansion}, it is possible to compute 
higher order corrections, whose coefficients are functions of the stereographic coordinates $\zeta_{a}$.
We obtain the structure
\begin{align}
\mathcal{M}_{5}\Bigg|_{\rm planar}&={3\zeta(3)\over 32}\alpha'^{3}\kappa^{3}s^{4}+
{5\zeta(5)\over 512}\alpha'^{5}\kappa^{3}s^{6}{Q_{10}(\zeta_{3},\zeta_{4},\zeta_{5})\over
(\zeta_{3}-\zeta_{4})^{2}(\zeta_{3}-\zeta_{5})^{2}(\zeta_{4}-\zeta_{5})^{2}} \nonumber \\[0.2cm]
&-{3\zeta(3)^{2}\over 2048}\alpha'^{6}\kappa^{3}s^{7}{Q_{12}(\zeta_{3},\zeta_{4},\zeta_{5})\over 
(\zeta_{3}-\zeta_{4})^{2}(\zeta_{3}-\zeta_{5})^{2}(\zeta_{4}-\zeta_{5})^{2}} \nonumber \\[0.2cm]
&+{7\zeta(7)\over 8192}\alpha'^{7}\kappa^{3}s^{8}{Q_{10}(\zeta_{3},\zeta_{4},\zeta_{5})^{2}\over 
(\zeta_{3}-\zeta_{4})^{4}(\zeta_{3}-\zeta_{5})^{4}(\zeta_{4}-\zeta_{5})^{4}} 
\label{eq:5grav_alpha'expansion} \\[0.2cm]
&-{\zeta(3)\zeta(5)\over 32768}\alpha'^{8}\kappa^{3}s^{9}{Q_{22}(\zeta_{3},\zeta_{4},\zeta_{5})\over
(\zeta_{3}-\zeta_{4})^{4}(\zeta_{3}-\zeta_{5})^{4}(\zeta_{4}-\zeta_{5})^{4}}+\ldots
\nonumber
\end{align}
The numerators $Q_{n}(\zeta_{3},\zeta_{4},\zeta_{5})$ appearing in this expansion are {\em nonhomogeneous} polynomials of 
degree $n$ whose explicit expressions are given in Eqs. \eqref{eq:Q10}-\eqref{eq:Q22} of the Appendix.
Our results show how the exchange of massive string modes renders the planar gravitational amplitude nonzero, with
the higher order terms in the $\alpha'$ expansion determined by nonhomogeneous polynomials. 

It is interesting to notice that the planar closed string amplitude \eqref{eq:5grav_alpha'expansion} does not
exhibit the soft poles at $\zeta_{a}\zeta_{b}=-1$ (with $a<b$), unlike the 
planar disk amplitude in Eq. \eqref{eq:disk_planar_amp}. This reflects the peculiar relation
between the soft and planar limits of amplitudes with gravitons, in both string and field theories.
It would be worthwhile to clarify the interplay between the two limits using recent results 
for soft theorems in string theory \cite{divecchia_et_al,sen}.

\section{Remarks on soft limits}
\label{sec:soft}

We turn now to the problem of whether the mathematical structure of planar zeros can be 
fully captured in the soft limit. We begin with the gauge case analyzing the
simple example of two distinguishable scalars studied in Section \ref{sec:disct_scalars_gauge}. In the limit in which the emitted positive
(resp. negative) helicity
gluon is soft, $p_{5}\rightarrow 0$, the leading behavior of the amplitude takes the form \cite{ours}
\begin{align}
\mathcal{A}_{5,\rm soft}&=2g\left(C_{1}{p_{3}\cdot\epsilon_{\pm}\over s_{35}}-C_{2}{p_{1}\cdot\epsilon_{\pm}\over s_{15}}+C_{4}{p_{4}\cdot\epsilon_{\pm}\over
s_{45}}-C_{5}{p_{2}\cdot\epsilon_{\pm}\over s_{25}}\right)\mathcal{A}_{4} \\[0.2cm]
&=2g\left[C_{1}\left({p_{3}\cdot\epsilon_{\pm}\over s_{35}}-{p_{2}\cdot\epsilon_{\pm}\over s_{25}}\right)
+C_{2}\left({p_{2}\cdot\epsilon_{\pm}\over s_{25}}-{p_{1}\cdot\epsilon_{\pm}\over s_{15}}\right)
+C_{4}\left({p_{4}\cdot\epsilon_{\pm}\over s_{45}}-{p_{2}\cdot\epsilon_{\pm}\over s_{25}}\right)\right]\mathcal{A}_{4},
\nonumber
\end{align}
where $\mathcal{A}_{4}$ is the four-scalar tree level amplitude. 
In terms of the stereographic coordinates $\zeta_{a}$ and taking the planar scattering limit, the 
soft amplitude reads 
\begin{align}
\mathcal{A}_{5,\rm soft}\Bigg|_{\rm planar}
&=\mp{g\sqrt{2}\over \sqrt{s}\zeta_{5}(1+\zeta_{3}\zeta_{4})} \nonumber \\[0.2cm]
&\times
\Big[(C_{1}-C_{2}+C_{4})\zeta_{3}\zeta_{4}
-(C_{1}-C_{2})\zeta_{3}\zeta_{5}+C_{2}-C_{4})\zeta_{4}\zeta_{5}-C_{2}\zeta_{5}^{2}\Big]\mathcal{A}_{4}.
\label{eq:gauge_amp_soft_limit}
\end{align}

The condition for the vanishing of the soft gauge theory amplitude in the planar limit is
given by
\begin{eqnarray}
(C_{1}-C_{2}+C_{4})\zeta_{3}\zeta_{4}
-(C_{1}-C_{2})\zeta_{3}\zeta_{5}+(C_{2}-C_{4})\zeta_{4}\zeta_{5}-C_{2}\zeta_{5}^{2}=0,
\end{eqnarray}
which reproduces the nontrivial loci of planar zeros for the full tree level amplitude discussed in Eq. \eqref{eq:condition_sQCD_2_full_zetas}. 
We notice, however, that in taking the soft limit we miss the trivial branch $2\zeta_{3}-\zeta_{4}=0$.  
In fact, this loci cannot be captured in the soft-gluon limit of the amplitude, since in the limit 
$\omega_{5}\rightarrow 0$,
\begin{eqnarray}
1+\zeta_{3}\zeta_{4}\longrightarrow 0 \hspace*{1cm} \Longrightarrow \hspace*{1cm} \zeta_{4}\longrightarrow -{1\over \zeta_{3}},
\end{eqnarray}
so we have
\begin{eqnarray}
2\zeta_{3}-\zeta_{4}\longrightarrow {2\zeta_{3}^{2}+1\over \zeta_{3}},
\end{eqnarray}
which implies that $2\zeta_{3}-\zeta_{4}$ never vanishes. 
This shows that the trivial branch of planar zeros is not accesible from 
the soft limit of the amplitude. Therefore, not all planar zeros can be realized in the limit in which the gluon is taken to be soft.
Notice, however, that this does not contradict the statements made in \cite{ours_pz}. Indeed, any planar zero can be realized in the 
limit in which {\em one} of the particles is taken to be soft. However, once we decide which particle is soft, not all planar zeros
can be realized in this regime, as we have seen in this case.

This being said, soft limits can be exploited to make a general analysis of planar zeros in the gauge case.
We study the scattering of $n$ charged particles in QED, parametrized by stereographic coordinates $\zeta_{i}$ 
($i=1,\ldots,n$), with the emission of a soft photon whose momenta we write in terms of the coordinate
$\zeta_{n+1}$,
\begin{eqnarray}
p_{a}&=&\omega_{a}\left(1,{\zeta_{a}+\overline{\zeta}_{a}\over 1+\zeta_{a}\overline{\zeta}_{a}},
i{\overline{\zeta}_{a}-\zeta_{a}\over 1+\zeta_{a}\overline{\zeta}_{a}},{\zeta_{a}\overline{\zeta}_{a}
-1\over 1+\zeta_{a}\overline{\zeta}_{a}}\right), \hspace*{1cm} a=1,\ldots,n+1.
\end{eqnarray}
The soft theorem for massless QED can be recast in terms of stereographic coordinates as \cite{he_et_al}
\begin{align}
\lim_{\omega_{n+1}\rightarrow 0^{+}}&\Big[\omega_{n+1}\mathcal{A}_{n+1}(p_{1},\dots,p_{n+1})\Big]
\nonumber \\[0.2cm]
&={1\over \sqrt{2}}(1+\zeta_{n+1}\overline{\zeta}_{n+1})\left(\sum_{i\in\rm out}{e_{i}\over \zeta_{i}
-\zeta_{n+1}}-\sum_{j\in\rm in}{e_{j}\over \zeta_{j}
-\zeta_{n+1}}\right)\mathcal{A}_{n}(p_{1},\ldots,p_{n}),
\end{align}
where we have used the following form for the polarization vector of the photon
\begin{eqnarray}
\epsilon_{+}={1\over \sqrt{2}}\Big(\overline{\zeta}_{n+1},1,-i,\overline{\zeta}_{n+1}\Big).
\end{eqnarray}
A planar zero is now obtained by setting
\begin{align}
\sum_{i\in\rm out}{e_{i}\over \zeta_{i}
-\zeta_{n+1}}=\sum_{j\in\rm in}{e_{j}\over \zeta_{j}
-\zeta_{n+1}},
\label{eq:planar_zero_nQED}
\end{align} 
with $\zeta_{1},\ldots,\zeta_{n+1}\in\mathbb{R}$. To compare with previous results, 
it is convenient to recast \eqref{eq:planar_zero_nQED} in the reference frame defined by 
Eqs. \eqref{eq:momentum_parametrization_in} and \eqref{eq:momentum_parametrization_out}. 
Setting $\zeta_{1}=\infty$ and $\zeta_{2}=0$, 
\begin{eqnarray}
\sum_{i=3}^{n}{e_{i}\over \zeta_{i}-\zeta_{n+1}}+{e_{2}\over \zeta_{n+1}}=0 \hspace*{0.5cm} \Longrightarrow \hspace*{0.5cm}
\zeta_{n+1}\sum_{i=3}^{n}e_{i}\prod_{i\neq\ell=3}^{n}(\zeta_{\ell}-\zeta_{n+1})
+e_{2}\prod_{\ell=3}^{n}(\zeta_{\ell}-\zeta_{n+1})=0.
\end{eqnarray}
The condition now is expressed in terms of a homogeneous polynomial of degree $n-2$ in the $n-1$ 
stereographic coordinates $(\zeta_{3},\ldots,\zeta_{n+1})$ parametrizing the momenta of the outgoing
particles. 
Particularizing the analysis to the five point amplitude and hard particles with charges $e_{1}=e_{4}=e$, $e_{2}=e_{3}=e'$, we have
\begin{eqnarray}
e'\zeta_{5}(\zeta_{3}-\zeta_{5})+e\zeta_{5}(\zeta_{4}-\zeta_{5})+e'(\zeta_{3}-\zeta_{5})(\zeta_{4}-\zeta_{5})
=0.
\end{eqnarray}
which is equivalent to \eqref{eq:maxwell_zeros} upon setting the projective coordinates defined in
\eqref{eq:projective_coordinates1}. 

\section{Concluding remarks}
\label{sec:closing}

It is indeed surprising that planar zeros of scattering amplitudes in (super) Yang-Mills
theories are determined by equations that are invariant under projective transformations of 
the stereographic coordinates associated with the directions of flight of the outgoing gauge bosons. 
In this paper we have shown that this is not a generic feature of field theories: while scalar fields
coupled to gauge bosons preserve the projective nature of planar zeros, pure scalar theories
have planar zeros that are not determined by projective curves. We have checked this explicitly 
in the case of the five-point amplitude in a theory of biadjoint scalars with cubic interactions. 

The projective nature of gauge planar zeros is also fragile with respect to the inclusion of string effects. 
We have seen how the $\alpha'$ corrections to the five gluon amplitude introduces terms which do not 
share the projective structure of the field theory result. 

The features of planar gravitational scattering differ in many aspects from those of gauge theories.
Due to the peculiar features of three-dimensional gravity, odd-multiplicity amplitudes are zero in the planar limit
while for even multiplicities they are only nonzero when helicity is conserved. We have checked this fact 
explicitly in various cases. String corrections to the field theory amplitude are generically nonvanishing in the 
planar limit, 
independently of their helicities and multiplicities, thus correcting the strong constraints imposed by the
results of \cite{huang_johansson}.

There are some intriguing elements in the interplay between planar zeros and soft limits in gauge theories that
are worth exploring. 
Although planar zeros are expected to be corrected by quantum effects, the very fact that they are determined
by the soft limit indicate that they might be of relevance for the infrared properties of the theory. 
In particular, it would be interesting to explore whether planar zeros are of any relevance for the asymptotic symmetries
for theories like QED \cite{he_et_al,QED_asymp,strominger_rev}.

\section*{Acknowledgments}

A.S.V. and D.M.J. acknowledge support from the Spanish Government grant FPA2015-65480-P 
and Spanish MINECO Centro de Excelencia Severo Ochoa Programme (SEV-2012-0249). 
The work of M.A.V.-M. has been partially supported by Spanish Government grant FPA2015-64041-C2-2-P. He also thanks
the Kavli Institute for the Physics and Mathematics of the Universe at the University of Tokyo for hospitality during 
the completion of this work.

\appendix

\section{Explicit expressions}
\label{sec:appendix}

\subsection{The numerator $\boldsymbol{P_{10}(\zeta_{3},\zeta_{4},\zeta_{5})}$ in equation 
\eqref{eq:scalar_amplitude_gen}}
\label{sec:appendixP10}

Here we give the explicit expression of the numerator $P_{10}(\zeta_{3},\zeta_{4},\zeta_{5})$ in 
Eq. \eqref{eq:scalar_amplitude_gen}, for generic color factors 
\begin{align}
P_{10}(\zeta_{3},\zeta_{4},\zeta_{5})&=c_{7}\overline{c}_{7}
\zeta_{3}^{6}\zeta_{4}^{3} \zeta_{5}+(c_{15}\overline{c}_{15}
-c_{7}\overline{c}_{7}-c_{8}\overline{c}_{8})\zeta_{3}^{6} \zeta_{4}^{2}\zeta_{5}^{2} 
     \nonumber \\[0.2cm]
&+c_{8}\overline{c}_{8}\zeta_{3}^{6}\zeta_{4}\zeta_{5}^{3} 
-c_{7}\overline{c}_{7}\zeta_{3}^{5}\zeta_{4}^{4}\zeta_{5}
+(2c_{8}\overline{c}_{8}-c_{7}\overline{c}_{7}-2c_{15}\overline{c}_{15})\zeta_{3}^{5}\zeta_{4}^{3}\zeta_{5}^{2}
    \nonumber \\[0.2cm]
&+c_{7}\overline{c}_{7}\zeta_{3}^{5}\zeta_{4}^{3}
+(2c_{7}\overline{c}_{7}-c_{8}\overline{c}_{8}-2c_{15}\overline{c}_{15})\zeta_{3}^{5}\zeta_{4}^{2}\zeta_{5}^{3}-c_{7}\overline{c}_{7}\zeta_{3}^{5}\zeta_{4}^{2}\zeta_{5} 
    \nonumber \\[0.2cm]
&-c_{8}\overline{c}_{8}\zeta_{3}^{5}\zeta_{4}\zeta_{5}^{4}
-c_{8}\overline{c}_{8}\zeta_{3}^{5}\zeta_{4}\zeta_{5}^{2}
+c_{8}\overline{c}_{8}\zeta_{3}^{5}\zeta_{5}^{3}-c_{6}\overline{c}_{6}\zeta_{3}^{4}\zeta_{4}^{5}\zeta_{5}
   \nonumber \\[0.2cm]
&+(c_{10}\overline{c}_{10}-c_{13}\overline{c}_{13}+c_{15}\overline{c}_{15}-
c_{3}\overline{c}_{3}+c_{5}\overline{c}_{5}+2c_{6}\overline{c}_{6}+2c_{7}\overline{c}_{7}
-c_{8}\overline{c}_{8})\zeta_{3}^{4}\zeta_{4}^{4}\zeta_{5}^{2} 
     \nonumber \\[0.2cm]
&+(c_{2}\overline{c}_{2}+c_{3}\overline{c}_{3}-c_{1}\overline{c}_{1}-2c_{10}\overline{c}_{10}-c_{11}\overline{c}_{11}
+c_{12}\overline{c}_{12}+c_{13}\overline{c}_{13}
     \nonumber \\[0.2cm]
&+4c_{15}\overline{c}_{15}-2c_{5}\overline{c}_{5}
-c_{6}\overline{c}_{6}-c_{7}\overline{c}_{7}-c_{8}\overline{c}_{8})\zeta_{3}^{4}\zeta_{4}^{3}\zeta_{5}^{3}
     \nonumber \\[0.2cm]
&+(c_{14}\overline{c}_{14}-c_{6}\overline{c}_{6}-c_{7}\overline{c}_{7})\zeta_{3}^{4}\zeta_{4}^{4}
+(2c_{6}\overline{c}_{6}-c_{7}\overline{c}_{7}-2c_{14}\overline{c}_{14})\zeta_{3}^{4}\zeta_{4}^{3}\zeta_{5}
     \nonumber \\[0.2cm]
&+(c_{10}\overline{c}_{10}+2c_{11}\overline{c}_{11}-c_{12}\overline{c}_{12}
+c_{15}\overline{c}_{15}-c_{2}\overline{c}_{2}+c_{5}\overline{c}_{5}-c_{7}\overline{c}_{7}
+2c_{8}\overline{c}_{8})\zeta_{3}^{4}\zeta_{4}^{2}\zeta_{5}^{4} 
     \nonumber \\[0.2cm]
&+(c_{4}\overline{c}_{4}-c_{1}\overline{c}_{1}-c_{11}\overline{c}_{11}+c_{14}\overline{c}_{14}
-c_{6}\overline{c}_{6}+2c_{7}\overline{c}_{7}+2c_{8}\overline{c}_{8}+c_{9}\overline{c}_{9})\zeta_{3}^{4}\zeta_{4}^{2}\zeta_{5}^{2}
     \label{eq:P10_general_biadjoing} \\[0.2cm]
&-c_{11}\overline{c}_{11}\zeta_{3}^{4}\zeta_{4}\zeta_{5}^{5}
+(2c_{11}\overline{c}_{11}-c_{8}\overline{c}_{8}-2c_{9}\overline{c}_{9})\zeta_{3}^{4}\zeta_{4}\zeta_{5}^{3}
+(c_{9}\overline{c}_{9}-c_{8}\overline{c}_{8}-c_{11}\overline{c}_{11})\zeta_{3}^{4}\zeta_{5}^{4}
     \nonumber \\[0.2cm]
&+c_{6}\overline{c}_{6}\zeta_{3}^{3}\zeta_{4}^{6}\zeta_{5}
+(2c_{13}\overline{c}_{13}-c_{6}\overline{c}_{6}-2c_{10}\overline{c}_{10})\zeta_{3}^{3}\zeta_{4}^{5}\zeta_{5}^{2}
+c_{6}\overline{c}_{6}\zeta_{3}^{3}\zeta_{4}^{5}
     \nonumber \\[0.2cm]
&+(c_{1}\overline{c}_{1}+4c_{10}\overline{c}_{10}+c_{11}\overline{c}_{11}-c_{12}\overline{c}_{12}
-c_{13}\overline{c}_{13}-2c_{15}\overline{c}_{15}-c_{2}\overline{c}_{2}+c_{3}\overline{c}_{3} 
     \nonumber \\[0.2cm] 
&-2c_{5}\overline{c}_{5}-c_{6}\overline{c}_{6}-c_{7}\overline{c}_{7}+c_{8}\overline{c}_{8})\zeta_{3}^{3} \zeta_{4}^{4}\zeta_{5}^{3}
+(2c_{7}\overline{c}_{7}-c_{6}\overline{c}_{6}-2c_{14}\overline{c}_{14})\zeta_{3}^{3}\zeta_{4}^{4}\zeta_{5}
     \nonumber \\[0.2cm]
&+(c_{1}\overline{c}_{1}-2c_{10}\overline{c}_{10}-c_{11}\overline{c}_{11}+c_{12}\overline{c}_{12}
-c_{13}\overline{c}_{13}-2c_{15}\overline{c}_{15}-c_{2}\overline{c}_{2}-c_{3}\overline{c}_{3} 
     \nonumber \\[0.2cm]
&+4c_{5}\overline{c}_{5}+c_{6}\overline{c}_{6}+c_{7}\overline{c}_{7}-c_{8}\overline{c}_{8})\zeta_{3}^{3}\zeta_{4}^{3}\zeta_{5}^{4}
+(c_{1}\overline{c}_{1}+c_{11}\overline{c}_{11}+c_{12}\overline{c}_{12}-c_{13}\overline{c}_{13}
     \nonumber \\[0.2cm] 
&+4c_{14}\overline{c}_{14}+c_{2}\overline{c}_{2}-c_{3}\overline{c}_{3}-2c_{4}\overline{c}_{4}
-c_{6}\overline{c}_{6}-c_{7}\overline{c}_{7}-c_{8}\overline{c}_{8}-2c_{9}\overline{c}_{9})
\zeta_{3}^{3}\zeta_{4}^{3}\zeta_{5}^{2} 
     \nonumber \\[0.2cm]
&+(2c_{2}\overline{c}_{2}-2c_{5}\overline{c}_{5}-c_{11}\overline{c}_{11})\zeta_{3}^{3}\zeta_{4}^{2}\zeta_{5}^{5}
+(c_{1}\overline{c}_{1}-c_{11}\overline{c}_{11}-c_{12}\overline{c}_{12}
+c_{13}\overline{c}_{13} 
     \nonumber \\[0.2cm] 
&-2c_{14}\overline{c}_{14}-c_{2}\overline{c}_{2}
+c_{3}\overline{c}_{3}-2c_{4}\overline{c}_{4}+c_{6}\overline{c}_{6}-c_{7}\overline{c}_{7}
-c_{8}\overline{c}_{8}+4c_{9}\overline{c}_{9})\zeta_{3}^{3}\zeta_{4}^{2}\zeta_{5}^{3} 
    \nonumber \\[0.2cm]
&+c_{11}\overline{c}_{11}\zeta_{3}^{3}\zeta_{4}\zeta_{5}^{6}
+(2c_{8}\overline{c}_{8}-2c_{9}\overline{c}_{9}-c_{11}\overline{c}_{11})\zeta_{3}^{3}\zeta_{4}\zeta_{5}^{4}
+c_{11}\overline{c}_{11}\zeta_{3}^{3}\zeta_{5}^{5}
      \nonumber \\[0.2cm]
&+(c_{10}\overline{c}_{10}-c_{6}\overline{c}_{6}-c_{13}\overline{c}_{13})\zeta_{3}^{2}\zeta_{4}^{6}\zeta_{5}^{2}
+(2c_{6}\overline{c}_{6}-2c_{10}\overline{c}_{10}-c_{13}\overline{c}_{13})\zeta_{3}^{2}\zeta_{4}^{5}\zeta_{5}^{3} 
      \nonumber \\[0.2cm]
&-c_{6}\overline{c}_{6}\zeta_{3}^{2}\zeta_{4}^{5}\zeta_{5}+(-c_{1}\overline{c}_{1}+c_{10}\overline{c}_{10}-c_{11}\overline{c}_{11}
+2c_{13}\overline{c}_{13}+c_{15}\overline{c}_{15} 
      \nonumber \\[0.2cm]
&+2c_{2}\overline{c}_{2}
+c_{5}\overline{c}_{5}-c_{6}\overline{c}_{6})\zeta_{3}^{2}\zeta_{4}^{4}\zeta_{5}^{4}
+(-c_{12}\overline{c}_{12}+2c_{13}\overline{c}_{13}+c_{14}\overline{c}_{14}-c_{2}\overline{c}_{2}
+c_{4}\overline{c}_{4} 
      \nonumber \\[0.2cm]
&+2c_{6}\overline{c}_{6}-c_{7}\overline{c}_{7}+c_{9}\overline{c}_{9})\zeta_{3}^{2}\zeta_{4}^{4}\zeta_{5}^{2}
+(-c_{2}\overline{c}_{2}-2c_{5}\overline{c}_{5}+2c_{11}\overline{c}_{11})\zeta_{3}^{2}\zeta_{4}^{3}\zeta_{5}^{5} 
      \nonumber \\[0.2cm]
&+(-c_{1}\overline{c}_{1}-c_{11}\overline{c}_{11}+c_{12}\overline{c}_{12}-
c_{13}\overline{c}_{13}-2c_{14}\overline{c}_{14}-c_{2}\overline{c}_{2}
+c_{3}\overline{c}_{3} 
      \nonumber \\[0.2cm]
&+4c_{4}\overline{c}_{4}-c_{6}\overline{c}_{6}+c_{7}\overline{c}_{7}
+c_{8}\overline{c}_{8}-2c_{9}\overline{c}_{9})\zeta_{3}^{2}\zeta_{4}^{3}\zeta_{5}^{3}
+(-c_{2}\overline{c}_{2}+c_{5}\overline{c}_{5}-c_{11}\overline{c}_{11})\zeta_{3}^{2}\zeta_{4}^{2}\zeta_{5}^{6}
      \nonumber \\[0.2cm]
&+(2c_{11}\overline{c}_{11}-c_{13}\overline{c}_{13}+c_{14}\overline{c}_{14}+2c_{2}\overline{c}_{2}
-c_{3}\overline{c}_{3}+c_{4}\overline{c}_{4}-c_{8}\overline{c}_{8}+c_{9}\overline{c}_{9})\zeta_{3}^{2}\zeta_{4}^{2}\zeta_{5}^{4} 
      \nonumber \\[0.2cm]
&-c_{11}\overline{c}_{11}\zeta_{3}^{2}\zeta_{4}\zeta_{5}^{5}
+c_{13}\overline{c}_{13}\zeta_{3}\zeta_{4}^{6}\zeta_{5}^{3}
-c_{13}\overline{c}_{13}\zeta_{3}\zeta_{4}^{5}\zeta_{5}^{4}
-c_{13}\overline{c}_{13}\zeta_{3}\zeta_{4}^{5}\zeta_{5}^{2}
-c_{2}\overline{c}_{2}\zeta_{3}\zeta_{4}^{4}\zeta_{5}^{5}
      \nonumber \\[0.2cm]
&+(2c_{2}\overline{c}_{2}-2c_{4}\overline{c}_{4}-c_{13}\overline{c}_{13})\zeta_{3}\zeta_{4}^{4}\zeta_{5}^{3}
+c_{2}\overline{c}_{2}\zeta_{3}\zeta_{4}^{3}\zeta_{5}^{6}
+(-c_{2}\overline{c}_{2}-2c_{4}\overline{c}_{4}+2c_{13}\overline{c}_{13})\zeta_{3}\zeta_{4}^{3}\zeta_{5}^{4}
     \nonumber \\[0.2cm]
&-c_{2}\overline{c}_{2}\zeta_{3}\zeta_{4}^{2}\zeta_{5}^{5}
+c_{13}\overline{c}_{13}\zeta_{4}^{5}\zeta_{5}^{3}
+(-c_{2}\overline{c}_{2}+c_{4}\overline{c}_{4}-c_{13}\overline{c}_{13})\zeta_{4}^{4}\zeta_{5}^{4}
+c_{2}\overline{c}_{2}\zeta_{4}^{3}\zeta_{5}^{5}.
     \nonumber
\end{align}
Due to Bose symmetry, the polynomial is invariant under permutations of its three 
variables $\zeta_{3}$, $\zeta_{4}$, and $\zeta_{5}$, provided this is supplemented with the
corresponding permutation of $S_{3}$ acting on the color factors, as explained in \cite{ours_pz}.

\subsection{The coefficients $\boldsymbol{A_{5}^{(2)}}$ and $\boldsymbol{A_{5}^{(3)}}$ of the $\boldsymbol{\alpha'}$ expansion \eqref{eq:disk_planar_amp}}
\label{sec:appendixA's}

The coefficient $A_{5}^{(2)}$ of the $\alpha'^{\,2}$ correction to the five-gluon amplitude is a degree 10 polynomial
in the stereographic coordinates, containing monomials of degree 8, 6, 4, and 2 as well
\begin{align}
A_{5}^{(2)}(\zeta_{3},\zeta_{4},\zeta_{5})&=-(c_{6}+c_{7})\zeta_{3}\zeta_{4}
-(c_{8}+c_{11})\zeta_{3}\zeta_{4}-(c_{2}+c_{13})\zeta_{4}\zeta_{5} \nonumber \\[0.2cm]
&-2(c_{6}+c_{7}+c_{8}+c_{11})\zeta_{3}^{2}\zeta_{4}\zeta_{5}-2(c_{2}+c_{6}+c_{7}+c_{13})
\zeta_{3}\zeta_{4}^{2}\zeta_{5} \nonumber \\[0.2cm]
&-2(c_{2}+c_{8}+c_{11}+c_{13})\zeta_{3}\zeta_{4}\zeta_{5}^{2}
-4(c_{2}+c_{6}+c_{7}+c_{8}+c_{11}+c_{13})\zeta_{3}^{2}\zeta_{4}^{2}\zeta_{5}^{2} \nonumber \\[0.2cm]
&+(c_{2}-c_{7}-c_{11}-c_{13})\zeta_{3}^{3}\zeta_{4}^{2}\zeta_{5}
+(-c_{7}+c_{8}-c_{11}-c_{13})\zeta_{3}\zeta_{4}^{3}\zeta_{5}^{2} \nonumber \\[0.2cm]
&+(-c_{2}-c_{6}-c_{8}+c_{11})\zeta_{3}^{2}\zeta_{4}^{3}\zeta_{5}
+(-c_{2}-c_{6}-c_{8}+c_{13})\zeta_{3}^{3}\zeta_{4}\zeta_{5}^{2} \label{eq:A_5^(2)} \\[0.2cm]
&+(c_{6}-c_{7}-c_{11}-c_{13})\zeta_{3}^{2}\zeta_{4}\zeta_{5}^{3}
+(-c_{2}-c_{6}+c_{7}-c_{8})\zeta_{3}\zeta_{4}^{2}\zeta_{5}^{3} \nonumber \\[0.2cm]
&-2(c_{6}+c_{7}+c_{8}+c_{13})\zeta_{3}^{3}\zeta_{4}^{3}\zeta_{5}^{2}
-2(c_{2}+c_{7}+c_{8}+c_{11})\zeta_{3}^{3}\zeta_{4}^{2}\zeta_{5}^{3} \nonumber \\[0.2cm]
&-2(c_{2}+c_{6}+c_{11}+c_{13})\zeta_{3}^{2}\zeta_{4}^{3}\zeta_{5}^{3} 
-(c_{7}+c_{8})\zeta_{3}^{4}\zeta_{4}^{3}\zeta_{5}^{3}-(c_{6}+c_{13})\zeta_{3}^{3}\zeta_{4}^{4}\zeta_{5}^{3}
\nonumber \\[0.2cm]
&-(c_{2}+c_{11})\zeta_{3}^{3}\zeta_{4}^{3}\zeta_{5}^{4}. \nonumber
\end{align}
The coefficient $A_{5}^{(3)}$ contains monomials of degree 15, 13, 11, 9, 7, 5, and 3
\begin{align}
A_{5}^{(3)}(\zeta_{3},\zeta_{4},\zeta_{5})&=(-c_{2}-c_{6}+c_{7}-c_{8}+c_{11}+c_{13})\zeta_{3}\zeta_{4}\zeta_{5}
-c_{7}\zeta_{3}^{2}\zeta_{4}+c_{6}\zeta_{3}\zeta_{4}^{2} \nonumber \\[0.2cm]
&+c_{8}\zeta_{3}^{2}\zeta_{5}-c_{11}\zeta_{3}\zeta_{5}^{2}-c_{13}\zeta_{4}^{2}\zeta_{5}+c_{2}\zeta_{4}\zeta_{5}^{2}
+3(-c_{7}+c_{8})\zeta_{3}^{3}\zeta_{4}\zeta_{5}+3(c_{6}-c_{13})\zeta_{3}\zeta_{4}^{3}\zeta_{5} \nonumber \\[0.2cm]
&+3(c_{2}-c_{11})\zeta_{3}\zeta_{4}\zeta_{5}^{3}
+3(-c_{2}+c_{6}-c_{7}-c_{8}+c_{11}+c_{13})\zeta_{3}^{2}\zeta_{4}^{2}\zeta_{5} \nonumber \\[0.2cm]
&+3(-c_{2}-c_{6}+c_{7}+c_{8}-c_{11}+c_{13})\zeta_{3}^{2}\zeta_{4}\zeta_{5}^{2}
+3(c_{2}-c_{6}+c_{7}-c_{8}+c_{11}-c_{13})\zeta_{3}\zeta_{4}^{2}\zeta_{5}^{2} \nonumber \\[0.2cm]
&+3(-c_{2}+c_{6}-c_{7}-c_{8}+c_{11}+c_{13})\zeta_{3}^{3}\zeta_{4}^{3}\zeta_{5}
+3(-c_{2}-c_{6}+c_{7}+c_{8}-c_{11}+c_{13})\zeta_{3}^{3}\zeta_{4}\zeta_{5}^{3} \nonumber \\[0.2cm]
&+3(c_{2}-c_{6}+c_{7}-c_{8}+c_{11}-c_{13})\zeta_{3}\zeta_{4}^{3}\zeta_{5}^{3} 
+6(-c_{2}+c_{6}-c_{7}+c_{11})\zeta_{3}^{2}\zeta_{4}^{2}\zeta_{5}^{3} \nonumber \\[0.2cm]
&+6(c_{6}-c_{8}+c_{11}-c_{13})\zeta_{3}^{2}\zeta_{4}^{3}\zeta_{5}^{2}
+6(-c_{2}-c_{7}+c_{8}+c_{13})\zeta_{3}^{3}\zeta_{4}^{2}\zeta_{5}^{2} \nonumber \\[0.2cm]
&-3c_{7}\zeta_{3}^{4}\zeta_{4}\zeta_{5}^{2}+3c_{6}\zeta_{3}\zeta_{4}^{4}\zeta_{5}^{2}
+3c_{8}\zeta_{3}^{4}\zeta_{4}^{2}\zeta_{5}-3c_{11}\zeta_{3}\zeta_{4}^{2}\zeta_{5}^{4}
-3c_{13}\zeta_{3}^{2}\zeta_{4}^{4}\zeta_{5}+3c_{2}\zeta_{3}^{2}\zeta_{4}\zeta_{5}^{4} \nonumber \\[0.2cm]
&+2(c_{6}-c_{7})\zeta_{3}^{4}\zeta_{4}^{4}\zeta_{5}+2(c_{8}-c_{11})\zeta_{3}^{4}\zeta_{4}\zeta_{5}^{4}
+2(c_{2}-c_{13})\zeta_{3}\zeta_{4}^{4}\zeta_{5}^{4} \nonumber \\[0.2cm]
&+2(-c_{7}+c_{8})\zeta_{3}^{5}\zeta_{4}^{2}\zeta_{5}^{2}+2(c_{6}-c_{13})\zeta_{3}^{2}\zeta_{4}^{5}\zeta_{5}^{2}
+2(c_{2}-c_{11})\zeta_{3}^{2}\zeta_{4}^{2}\zeta_{5}^{5} \nonumber \\[0.2cm]
&+3(c_{2}+c_{6}-c_{7}+c_{8}-c_{11}-c_{13})\zeta_{3}^{3}\zeta_{4}^{3}\zeta_{5}^{3} \nonumber \\[0.2cm]
&+(-5c_{2}+3c_{6}-3c_{7}+3c_{8}+5c_{11}+5c_{13})\zeta_{3}^{4}\zeta_{4}^{3}\zeta_{5}^{2} 
\label{eq:A_5^(3)} \\[0.2cm]
&+(-5c_{2}+3c_{6}-3c_{7}-5c_{8}+5c_{11}-3c_{13})\zeta_{3}^{3}\zeta_{4}^{4}\zeta_{5}^{2} \nonumber \\[0.2cm]
&+(-5c_{2}-5c_{6}-3c_{7}+3c_{8}-3c_{11}+5c_{13})\zeta_{3}^{4}\zeta_{4}^{2}\zeta_{5}^{3} \nonumber \\[0.2cm]
&+(3c_{2}+3c_{6}+5c_{7}-5c_{8}+5c_{11}-3c_{13})\zeta_{3}^{2}\zeta_{4}^{4}\zeta_{5}^{3} \nonumber \\[0.2cm]
&+(3c_{2}-5c_{6}+5c_{7}+3c_{8}-3c_{11}+5c_{13})\zeta_{3}^{3}\zeta_{4}^{2}\zeta_{5}^{4} \nonumber \\[0.2cm]
&+(3c_{2}-5c_{6}+5c_{7}-5c_{8}-3c_{11}-3c_{13})\zeta_{3}^{2}\zeta_{4}^{3}\zeta_{5}^{4} \nonumber \\[0.2cm]
&+3(-c_{2}-c_{6}-c_{7}+c_{8}+c_{11}+c_{13})\zeta_{3}^{5}\zeta_{4}^{3}\zeta_{5}^{3} 
+3(-c_{2}+c_{6}+c_{7}-c_{8}+c_{11}-c_{13})\zeta_{3}^{3}\zeta_{4}^{5}\zeta_{5}^{3} \nonumber \\[0.2cm]
&+3(c_{2}-c_{6}+c_{7}-c_{8}-c_{11}+c_{13})\zeta_{3}^{3}\zeta_{4}^{3}\zeta_{5}^{5} 
+6(-c_{2}+c_{6}-c_{7}+c_{11})\zeta_{3}^{4}\zeta_{4}^{4}\zeta_{5}^{3} \nonumber \\[0.2cm]
&+6(-c_{6}+c_{8}-c_{11}+c_{13})\zeta_{3}^{4}\zeta_{4}^{3}\zeta_{5}^{4}
+6(c_{2}+c_{7}-c_{8}-c_{13})\zeta_{3}^{3}\zeta_{4}^{4}\zeta_{5}^{4} \nonumber \\[0.2cm]
&+3(c_{6}-c_{7})\zeta_{3}^{5}\zeta_{4}^{5}\zeta_{5}^{3}+3(c_{8}-c_{11})\zeta_{3}^{5}\zeta_{4}^{3}\zeta_{5}^{5}
+3(c_{2}-c_{13})\zeta_{3}^{3}\zeta_{4}^{5}\zeta_{5}^{5} \nonumber \\[0.2cm]
&+3(-c_{2}-c_{6}-c_{7}+c_{8}+c_{11}+c_{13})\zeta_{3}^{5}\zeta_{4}^{4}\zeta_{5}^{4}
+(-c_{2}+c_{6}+c_{7}-c_{8}+c_{11}-c_{13})\zeta_{3}^{4}\zeta_{4}^{5}\zeta_{5}^{4} \nonumber \\[0.2cm]
&+3(c_{2}-c_{6}+c_{7}-c_{8}-c_{11}+c_{13})\zeta_{3}^{4}\zeta_{4}^{4}\zeta_{5}^{5}
-c_{7}\zeta_{3}^{6}\zeta_{4}^{5}\zeta_{5}^{4}+c_{6}\zeta_{3}^{5}\zeta_{4}^{6}\zeta_{5}^{4} \nonumber \\[0.2cm]
&+c_{8}\zeta_{3}^{6}\zeta_{4}^{4}\zeta_{5}^{5}-c_{11}\zeta_{3}^{4}\zeta_{4}^{4}\zeta_{5}^{6}
-c_{13}\zeta_{3}^{4}\zeta_{4}^{6}\zeta_{5}^{5}+c_{2}\zeta_{3}^{4}\zeta_{4}^{5}\zeta_{5}^{6}
\nonumber \\[0.2cm]
&+(-c_{2}-c_{6}+c_{7}-c_{8}+c_{11}+c_{13})\zeta_{3}^{5}\zeta_{4}^{5}\zeta_{5}^{5}.
\nonumber
\end{align}

\subsection{Numerators of the $\boldsymbol{\alpha'}$ corrections to the gravitational planar amplitude}

The $\alpha'$ expansion of the string graviton amplitude is given in Eq. \eqref{eq:5grav_alpha'expansion}. 
The ten-degree polynomial appearing in both the $\alpha'^{5}$ and $\alpha'^{7}$ terms is given by
\begin{align}
Q_{10}(\zeta_{3},\zeta_{4},\zeta_{5})&=
\zeta_{4}^{2}\zeta_{5}^{4}\zeta_{3}^{4}-\zeta_{4}^{3}\zeta_{5}^{3}\zeta_{3}^{4}
+\zeta_{4}\zeta_{5}^{3}\zeta_{3}^{4}+\zeta_{4}^{2}\zeta_{3}^{4}+\zeta_{4}^{4}\zeta_{5}^{2}\zeta_{3}^{4}
-\zeta_{4}^{2}\zeta_{5}^{2}\zeta_{3}^{4}  \nonumber \\[0.2cm]
&+\zeta_{5}^{2}\zeta_{3}^{4}+\zeta_{4}^{3}\zeta_{5}\zeta_{3}^{4}-\zeta_{4}\zeta_{5}\zeta_{3}^{4}
-\zeta_{4}^{3}\zeta_{5}^{4}\zeta_{3}^{3}+\zeta_{4}\zeta_{5}^{4}\zeta_{3}^{3}-\zeta_{4}^{3}\zeta_{3}^{3} 
\nonumber \\[0.2cm]
&-\zeta_{4}^{4}\zeta_{5}^{3}\zeta_{3}^{3}-\zeta_{4}^{2}\zeta_{5}^{3}\zeta_{3}^{3}-\zeta_{5}^{3}\zeta_{3}^{3}-\zeta_{4}^{3}\zeta_{5}^{2}\zeta_{3}^{3}
+\zeta_{4}\zeta_{5}^{2}\zeta_{3}^{3}+\zeta_{4}\zeta_{3}^{3} \nonumber \\[0.2cm]
&+\zeta_{4}^{4}\zeta_{5}\zeta_{3}^{3}+\zeta_{4}^{2}\zeta_{5}\zeta_{3}^{3}
+\zeta_{5}\zeta_{3}^{3}+\zeta_{4}^{4}\zeta_{3}^{2}+\zeta_{4}^{4}\zeta_{5}^{4}\zeta_{3}^{2}-\zeta_{4}^{2}\zeta_{5}^{4}\zeta_{3}^{2} \nonumber \\[0.2cm]
&+\zeta_{5}^{4}\zeta_{3}^{2}-\zeta_{4}^{3}\zeta_{5}^{3}\zeta_{3}^{2}+\zeta_{4}\zeta_{5}^{3}\zeta_{3}^{2}-\zeta_{4}^{2}\zeta_{3}^{2}
-\zeta_{4}^{4}\zeta_{5}^{2}\zeta_{3}^{2}-6\zeta_{4}^{2}\zeta_{5}^{2}\zeta_{3}^{2}
\label{eq:Q10} \\[0.2cm]
&-\zeta_{5}^{2}\zeta_{3}^{2}+\zeta_{4}^{3}\zeta_{5}\zeta_{3}^{2}-\zeta_{4}\zeta_{5}\zeta_{3}^{2}
+\zeta_{3}^{2}+\zeta_{4}^{3}\zeta_{5}^{4}\zeta_{3}-\zeta_{4}\zeta_{5}^{4}\zeta_{3}+\zeta_{4}^{3}\zeta_{3} 
\nonumber \\[0.2cm]
&+\zeta_{4}^{4}\zeta_{5}^{3}\zeta_{3}+\zeta_{4}^{2}\zeta_{5}^{3}\zeta_{3}+\zeta_{5}^{3}\zeta_{3}
+\zeta_{4}^{3}\zeta_{5}^{2}\zeta_{3}-\zeta_{4}\zeta_{5}^{2}\zeta_{3}-\zeta_{4}\zeta_{3}  \nonumber \\[0.2cm]
&-\zeta_{4}^{4}\zeta_{5}\zeta_{3}-\zeta_{4}^{2}\zeta_{5}\zeta_{3}-\zeta_{5}\zeta_{3}+\zeta_{4}^{2}\zeta_{5}^{4}
-\zeta_{4}^{3}\zeta_{5}^{3}+\zeta_{4}\zeta_{5}^{3} \nonumber \\[0.2cm]
&+\zeta_{4}^{2}+\zeta_{4}^{4}\zeta_{5}^{2}-\zeta_{4}^{2}\zeta_{5}^{2}+\zeta_{5}^{2}+\zeta_{4}^{3}\zeta_{5}-\zeta_{4}\zeta_{5}.
\nonumber
\end{align}
It contains terms of degree 10, 8, 6, 4, and 2. The numerator associated with the $\alpha'^{6}$ term is 
\begin{align}
Q_{12}(\zeta_{3},\zeta_{4},\zeta_{5})&=2\zeta_{4}^{4}\zeta_{5}^{4}\zeta_{3}^{4}+3\zeta_{4}^{3}\zeta_{5}^{3}\zeta_{3}^{4}
+3\zeta_{4}^{2}\zeta_{5}^{2}\zeta_{3}^{4}+3\zeta_{4}^{3}\zeta_{5}^{4}\zeta_{3}^{3}
+3\zeta_{4}^{4}\zeta_{5}^{3}\zeta_{3}^{3}+3\zeta_{4}^{2}\zeta_{5}^{3}\zeta_{3}^{3} \nonumber \\[0.2cm]
&+3\zeta_{4}^{3}\zeta_{5}^{2}\zeta_{3}^{3}+4\zeta_{4}\zeta_{5}^{2}\zeta_{3}^{3}
+4\zeta_{4}^{2}\zeta_{5}\zeta_{3}^{3}+3\zeta_{4}^{2}\zeta_{5}^{4}\zeta_{3}^{2}
+3\zeta_{4}^{3}\zeta_{5}^{3}\zeta_{3}^{2}+4\zeta_{4}\zeta_{5}^{3}\zeta_{3}^{2} \nonumber \\[0.2cm]
&+3\zeta_{4}^{2}\zeta_{3}^{2}+3\zeta_{4}^{4}\zeta_{5}^{2}\zeta_{3}^{2}
-2\zeta_{4}^{2}\zeta_{5}^{2}\zeta_{3}^{2}+3\zeta_{5}^{2}\zeta_{3}^{2}
+4\zeta_{4}^{3}\zeta_{5}\zeta_{3}^{2}+3\zeta_{4}\zeta_{5}\zeta_{3}^{2}
\label{eq:Q12} \\[0.2cm]
&+4\zeta_{4}^{2}\zeta_{5}^{3}\zeta_{3}+4\zeta_{4}^{3}\zeta_{5}^{2}\zeta_{3}+3\zeta_{4}\zeta_{5}^{2}\zeta_{3}
+3\zeta_{4}\zeta_{3}+3\zeta_{4}^{2}\zeta_{5}\zeta_{3}+3\zeta_{5}\zeta_{3} \nonumber \\[0.2cm]
&+3\zeta_{4}^{2}\zeta_{5}^{2}+3\zeta_{4}\zeta_{5}+2.
\nonumber
\end{align}
This is a degree 12 polynomial including monomials of degree 12, 10, 8, 6, 4, 2, and 0. 
Finally, the numerator determining the $\alpha'^{8}$ corrections is the following nonhomogeneous degree 22 polynomial
\begin{align}
Q_{22}(\zeta_{3},\zeta_{4},\zeta_{5})&=8Q_{10}Q_{12}+3(1+\zeta_{3}\zeta_{4})^{2}(1+\zeta_{3}\zeta_{5})^{2}(1+\zeta_{4}\zeta_{5})^{2} 
\Big(2\zeta_{4}^{2}\zeta_{3}^{4}+2\zeta_{5}^{2}\zeta_{3}^{4} \nonumber \\[0.2cm]
&-2\zeta_{4}\zeta_{5}\zeta_{3}^{4}-2\zeta_{4}^{3}\zeta_{3}^{3}
-2\zeta_{5}^{3}\zeta_{3}^{3}-\zeta_{4}\zeta_{5}^{2}\zeta_{3}^{3}
-\zeta_{4}^{2}\zeta_{5}\zeta_{3}^{3}+2\zeta_{4}^{4}\zeta_{3}^{2}+2\zeta_{5}^{4}\zeta_{3}^{2} \nonumber \\[0.2cm]
&-\zeta_{4}\zeta_{5}^{3}\zeta_{3}^{2}+6\zeta_{4}^{2}\zeta_{5}^{2}\zeta_{3}^{2}
-\zeta_{4}^{3}\zeta_{5}\zeta_{3}^{2}-2\zeta_{4}\zeta_{5}^{4}\zeta_{3}-\zeta_{4}^{2}\zeta_{5}^{3}\zeta_{3}
-\zeta_{4}^{3}\zeta_{5}^{2}\zeta_{3}  \label{eq:Q22} \\[0.2cm]
&-2\zeta_{4}^{4}\zeta_{5}\zeta_{3}+2\zeta_{4}^{2}\zeta_{5}^{4}-2\zeta_{4}^{3}\zeta_{5}^{3}+2\zeta_{4}^{4}\zeta_{5}^{2}\Big),
\nonumber
\end{align}
where $Q_{10}$ and $Q_{12}$ are the polynomials given in Eqs. \eqref{eq:Q10} and \eqref{eq:Q12}.

\end{document}